\documentclass[%
 reprint,
superscriptaddress,
bibnotes,
 amsmath,amssymb,
 aps,
nobalancelastpage]{revtex4-1}

\usepackage{graphicx}
\usepackage{dcolumn}
\usepackage{bm}
\usepackage[many]{tcolorbox}
\usepackage{lipsum}
\usepackage{amsmath}
\usepackage{shuffle}
\usepackage{tikz}
\usepackage{environ}
\usepackage{tabularx}
\usetikzlibrary{shapes.geometric, arrows,decorations}
\usepackage{amsmath}
\usepackage{mathrsfs}
\usepackage{tikz}
\usepackage{tikz-3dplot}
\usepackage{graphicx}
\usepackage{subcaption}
\usepackage{overpic}
\usepackage{amsfonts}
\usepackage{float}
\usetikzlibrary{snakes}
\usetikzlibrary{calc}
\usetikzlibrary{decorations}

\usetikzlibrary{decorations.markings}

\makeatletter
\newsavebox{\measure@tikzpicture}
\NewEnviron{scaletikzpicturetowidth}[1]{%
  \def\tikz@width{#1}%
  \begin{lrbox}{\measure@tikzpicture}%
  \BODY
  \end{lrbox}%
  \pgfmathparse{#1/\wd\measure@tikzpicture}%
  \BODY
}
\makeatother
\begin{document}

\title{The symbology of Feynman integrals from twistor geometries}
\author{Song He}
\email{songhe@itp.ac.cn}
\affiliation{%
CAS Key Laboratory of Theoretical Physics, Institute of Theoretical Physics, Chinese Academy of Sciences, Beijing 100190, China
}%
\affiliation{
School of Fundamental Physics and Mathematical Sciences, Hangzhou Institute for Advanced Study, UCAS, Hangzhou 310024, China}
\affiliation{International Centre for Theoretical Physics Asia-Pacific, Beijing/Hangzhou, China}
\affiliation{%
Peng Huanwu Center for Fundamental Theory, Hefei, Anhui 230026, P. R. China
}%
\author{Jiahao Liu}
\email{liujiahao222@mails.ucas.ac.cn}
 \affiliation{%
CAS Key Laboratory of Theoretical Physics, Institute of Theoretical Physics, Chinese Academy of Sciences, Beijing 100190, China
}%

\affiliation{%
School of Physical Sciences, UCAS, 
No.19A Yuquan Road, Beijing 100049, China
}
\author{Yichao Tang}
\email{tangyichao@itp.ac.cn}
 \affiliation{%
CAS Key Laboratory of Theoretical Physics, Institute of Theoretical Physics, Chinese Academy of Sciences, Beijing 100190, China
}%

\affiliation{%
School of Physical Sciences, UCAS, 
No.19A Yuquan Road, Beijing 100049, China
}
 \author{Qinglin Yang}%
 \email{yangqinglin@itp.ac.cn}
 \affiliation{%
CAS Key Laboratory of Theoretical Physics, Institute of Theoretical Physics, Chinese Academy of Sciences, Beijing 100190, China
}%
\affiliation{%
School of Physical Sciences, UCAS, 
No.19A Yuquan Road, Beijing 100049, China
}

\date{\today}
\begin{abstract}
We study the symbology of planar Feynman integrals in dimensional regularization by considering geometric configurations in momentum twistor space corresponding to their leading singularities (LS). Cutting propagators in momentum twistor space amounts to intersecting lines associated with loop and external dual momenta, including the special line associated with the point at infinity, which breaks dual conformal symmetry. We show that cross-ratios of intersection points on these lines, especially those on the infinity line, naturally produce symbol letters for Feynman integrals in $D=4-2 \epsilon$, which include and generalize their LS. At one loop, we obtain all symbol letters using intersection points from quadruple cuts for integrals up to pentagon kinematics with two massive corners, which agree perfectly with canonical differential equation (CDE) results. We then obtain all two-loop letters, for up to four-mass box and one-mass pentagon kinematics, by considering more intersections arising from two-loop cuts. Finally we comment on how cluster algebras appear from this construction, and importantly how we may extend the method to non-planar integrals. 
\end{abstract}
\maketitle

\section{Introduction}

Since the inception of twistor-string theory in ${\cal N}=4$ supersymmetric Yang-Mills theory (SYM)~\cite{Witten:2003nn}, we have witnessed tremendous progress in computing and understanding the hidden structures of scattering amplitudes. In this endeavor, geometric considerations in terms of twistors and momentum twistors~\cite{Hodges:2009hk} have played a crucial role both for triggering powerful computational methods such as BCF(W) and CSW~\cite{Britto:2004ap,*Britto:2005fq, Cachazo:2004kj}, and for unravelling new mathematical structures such as positive Grassmannians and the amplituhedron~\cite{Arkani-Hamed:2016byb, Arkani-Hamed:2013jha}. Largely inspired by these developments, ${\cal N}=4$ SYM has also become an extremely fruitful laboratory for new methods of evaluating multi-loop Feynman integrals, which is a subject of enormous interest by itself. Significant progress has been made in computing dual conformal invariant (DCI) Feynman integrals contributing to amplitudes, most naturally formulated in momentum twistor space~\cite{Drummond:2006rz,*Drummond:2007aua,Drummond:2010cz,ArkaniHamed:2010gh,Spradlin:2011wp,DelDuca:2011wh,Bourjaily:2013mma,Henn:2018cdp,Herrmann:2019upk, Bourjaily:2018aeq,Bourjaily:2019hmc,He:2020uxy,He:2020lcu}. Remarkably, just like the full (integrated) amplitudes, individual Feynman integrals exhibit unexpected mathematical structures such as \emph{cluster algebras}~\cite{fomin2002cluster,*fomin2003cluster,*fomin2007cluster}: when integrals evaluate to multiple polylogarithm (MPL) functions, the \emph{symbol alphabets}~\cite{Goncharov:2010jf,Duhr:2011zq} organize into cluster algebras, {\it e.g.}, $A_3$ and $E_6$ for six- and seven-point amplitudes/integrals in SYM~\cite{Golden:2013xva,*Golden:2014xqa}. These alphabets are the starting point of bootstrapping such amplitudes to impressively high loop orders~\cite{Dixon:2011pw,*Dixon:2014xca,*Dixon:2014iba,*Dixon:2015iva,*Caron-Huot:2016owq,*Caron-Huot:2019vjl,*Caron-Huot:2019bsq, Drummond:2014ffa,*Dixon:2016nkn,*Drummond:2018caf,*Dixon:2020cnr, Golden:2021ggj, Caron-Huot:2020bkp}. It is highly non-trivial that cluster algebras appear for individual (all-loop) Feynman integrals beyond the $n=6,7$ cases~\cite{Caron-Huot:2018dsv, Drummond:2017ssj, He:2021esx,*He:2021non,*He:2021eec}.

It is natural to ask how much of these mathematical structures carry over to general, non-DCI Feynman integrals and in turn to amplitudes in realistic theories such as QCD. Recently, cluster algebraic structures have been identified and explored for the symbology of Feynman integrals in dimensional regularization~\cite{Chicherin:2020umh}, which also utilizes the power of momentum twistors in parametrizing kinematics. In this letter, we propose a new method to study the symbology of Feynman integrals in $D=4-2 \epsilon$, where \emph{twistor geometries} lie at the heart. Very recently, the so-called Schubert problems, which studies geometric configurations in momentum twistor space~\cite{Hodges:2010kq,ArkaniHamed:2010gh} for \emph{leading singularities} (LS)~\cite{Bern:1994zx,*Bern:1994cg,Britto:2004nc,Cachazo:2008vp} of DCI integrals in $D=4$, have proved very successful in predicting their symbol alphabets~\cite{Yang:2022gko,He:2022ctv}, {\it e.g.}, the $9+9$ algebraic letters for $n=8$ and various higher-point examples with massive legs~\cite{Zhang:2019vnm,He:2020lcu,He:2021fwf}. Note that one can always represent any massive momentum using two (or more) massless ones: {\it e.g.}, for the two-mass-easy box kinematics with $P_1^2, P_3^2\neq 0$ (and $P_2^2=P_4^2=0)$, we can introduce $6$ massless ones, with $p_6+p_1=P_1, p_2=P_2, p_3+p_4=P_3$ and $p_5=P_4$. Thus it suffices to consider planar kinematics with $n$ massless momenta, and we introduce dual momenta via $p_i^\mu=:x^\mu_{i{+}1}-x^\mu_i$ for $i=1,\cdots, n$ ($x_{n+1}\equiv x_1$) satisfying $x_{i,i{+}1}^2=0$. The two-mass-easy box kinematics thus depend on the dual momenta $\{x_2, x_3, x_5, x_6\}$. 

Such a planar massless kinematics can be described by $n$ {\it momentum twistors}~\cite{Hodges:2009hk}, $Z^{a=1,\cdots,4}_i\in\mathbb{P}^3$ satisfying projectivity $Z_i^a\sim t_i Z^a_i$ for each massless momentum. DCI quantities are built from homogeneous degree-$0$ functions, {\it e.g.}, cross-ratios of $\langle i j k l\rangle:={\rm det}(Z_i Z_j Z_k Z_l)$. 
The dual momentum $x_i^\mu$ corresponds to a line (bi-twistor) $(Z_{i{-}1}, Z_i)$, and $x_{i,j}^2=\frac{\langle i{-}1 i j{-}1 j\rangle}{\langle i{-}1 i I_\infty\rangle \langle j{-}1 j I_\infty\rangle}$. The factors involving the special point at infinity, $I_{\infty}$ (or infinity line), drop out in DCI quantities. In general, non-DCI planar integrals are naturally written in terms of momentum twistors, depending on not only the external lines for dual momenta, but also $I_\infty$ (see~\cite{Arkani-Hamed:2010pyv}). What we will show is that their symbology in $D=4-2 \epsilon$ are essentially identical to the DCI case dictated by twistor geometries, once we take this special infinity line into account.

Specifically, a planar loop integral has dual loop momenta, and each of them (say $x_0$) is represented by a line $(AB)$ in momentum twistor space. Strictly in $D=4$, we say the integral is DCI if it is independent of $I_{\infty}$. For example, for a box integral $I_4(i,j,k,l)$ with four dual momenta $\{x_i, x_j, x_k, x_l\}$, the propagators in terms of momentum twistors read $\frac 1 {(x_0-x_i)^2}=\frac{\langle AB I_{\infty}\rangle \langle i{-}1 i I_{\infty} \rangle}{\langle AB i{-}1 i\rangle}$, and the measure in $D=4$ is $d^4 x_0=\frac{d^4 A d^4 B}{{\rm vol. GL}(2) \langle AB I_{\infty}\rangle^4}$, thus all factors involving $\langle AB I_{\infty}\rangle$ cancel out and we are left with four inverse propagators of momentum twistors: $\langle AB i{-}1 i\rangle\langle AB j{-}1 j\rangle\langle AB k{-}1 k\rangle\langle AB l{-}1 l\rangle$~\footnote{It is well known that constant factors $\langle i{-}1 i I_{\infty}\rangle$ {\it etc.} cancel out by normalizing the box in terms of its LS.}. On the other hand, even in $D=4$, triangle integrals $I_3(i,j,k)$ depending on $\{x_i,x_j,x_k\}$ are not DCI; we have 3 inverse propagators and $\langle AB I_{\infty}\rangle$ in the denominator. For integrals in $D=4-2\epsilon$, DCI is always broken, which schematically means the presence of $\langle AB I_{\infty}\rangle$ with powers depending on $\epsilon$. In the following sections, we will present a method for producing symbol letters of Feynman integrals in $D=4-2 \epsilon$ as cross-ratios of intersection points on various lines, which are associated with their LS in momentum twistor space.

\section{Twistor geometries and symbology for one-loop integrals}
\begin{figure}[htbp]
    \centering
    \begin{subfigure}{0.4\linewidth}
        \centering
        \begin{tikzpicture}[scale=0.2]
            \draw[black,thick] (0,4)--(-3.46,2)--(0,0)--cycle;
            \draw[black,thick] (0,4)--(1,5.73);
            \draw[black,thick] (1.73,-1)--(0,0)--(0,-2);
            \filldraw[black] (1,5.73) node[anchor=south west] {{$i$}};
            \filldraw[black] (1.73,-1) node[anchor=west] {{$i{+}1$}};
            \filldraw[black] (0,-2) node[anchor=north] {{$i{-}2$}};
            \draw[black,thick] (-5.46,2)--(-3.46,2);
            \filldraw[black] (-5.46,2)node[anchor=east]{{$P_3\ i{-}1$}};
            \filldraw[black] (3,-3.46) node[anchor=center] {{$P_2$}};
            \filldraw[black] (4,5.73) node[anchor=south] {{$P_1$}};
        \end{tikzpicture}
        \caption{}
        \label{triangle1}
    \end{subfigure}
    \begin{subfigure}{0.4\linewidth}
        \centering
        \begin{tikzpicture}[scale=0.2]
            \draw[black,thick] (0,4)--(-3.46,2)--(0,0)--cycle;
            \draw[black,thick] (0,6)--(0,4)--(1.73,5);
            \draw[black,thick] (1.73,-1)--(0,0)--(0,-2);
            \filldraw[black] (0,6) node[anchor=south] {{$i{+}1$}};
            \filldraw[black] (1.73,5) node[anchor=west] {{$j{-}1$}};
            \filldraw[black] (1.73,-1) node[anchor=west] {{$j$}};
            \filldraw[black] (0,-2) node[anchor=north] {{$i{-}1$}};
            \draw[black,thick] (-5.46,2)--(-3.46,2);
            \filldraw[black] (-5.46,2)node[anchor=east]{{$P_3\ i$}};
            \filldraw[black] (3,-3.46) node[anchor=center] {{$P_2$}};
            \filldraw[black] (3,7.46) node[anchor=center] {{$P_1$}};
        \end{tikzpicture}
        \caption{}
        \label{triangle2}
    \end{subfigure}
    
    \begin{subfigure}{0.95\linewidth}
        \centering
        \begin{tikzpicture}[scale=0.2,baseline=(current bounding box.center)]
            \draw[black,thick] (0,4)--(-3.46,2)--(0,0)--cycle;
            \draw[black,thick] (0,6)--(0,4)--(1.73,5);
            \draw[black,thick] (1.73,-1)--(0,0)--(0,-2);
            \draw[black,thick] (-5.19,1)--(-3.46,2)--(-5.19,3);
            \filldraw[black] (0,6) node[anchor=south] {{$i$}};
            \filldraw[black] (1.73,5) node[anchor=west] {{$j{-}1$}};
            \filldraw[black] (1.73,-1) node[anchor=west] {{$j$}};
            \filldraw[black] (0,-2) node[anchor=north] {{$k{-}1$}};
            \filldraw[black] (-5.19,3) node[anchor=south east] {{$i{-1}$}};
            \filldraw[black] (-5.19,1) node[anchor=north east] {{$k$}};
            \filldraw[black] (3,7.46) node[anchor=center] {{$P_1$}};
            \filldraw[black] (3,-3.46) node[anchor=center] {{$P_2$}};
            \filldraw[black] (-7.46,2) node[anchor=center] {{$P_3$}};
        \end{tikzpicture}\hspace{-1.2em}
        $\xleftarrow{x_l\to I_\infty}$\hspace{-1.2em}
        \begin{tikzpicture}[scale=0.2,baseline=(current bounding box.center)]
            \draw[black,thick] (15,0)--(23,0);
            \draw[black,thick] (17,-2)--(17,6);
            \draw[black,thick] (15,4)--(23,4);
            \draw[black,thick] (21,-2)--(21,6);
            \filldraw[black] (15,0) node[anchor=east] {{$l{-}1$}};
            \filldraw[black] (15,4) node[anchor=east] {{$l$}};
            \filldraw[black] (17,6) node[anchor=south] {{$i{-}1$}};
            \filldraw[black] (21,6) node[anchor=south] {{$i$}};
            \filldraw[black] (23,4) node[anchor=west] {{$j{-}1$}};
            \filldraw[black] (23,0) node[anchor=west] {{$j$}};
            \filldraw[black] (21,-2) node[anchor=north] {{$k{-}1$}};
            \filldraw[black] (17,-2) node[anchor=north] {{$k$}};
        \end{tikzpicture}
        \caption{}
        \label{triangle3}
    \end{subfigure}
    \caption{One-mass \subref{triangle1}, two-mass \subref{triangle2} and three-mass \subref{triangle3} triangles.}
    \label{triangles}
\end{figure}
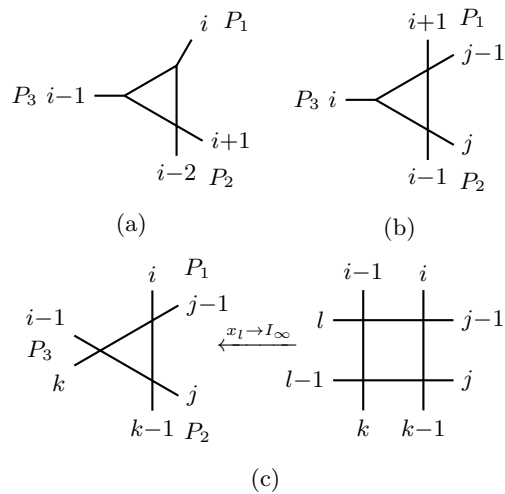

In this section, we show that symbol letters for one-loop integrals in $D=4- 2\epsilon$ can be obtained from geometric configurations in twistor space. We begin with triangle integrals, which turn out to be the building blocks of our one-loop constructions. Consider the one-mass triangle $I_3(i{-}1,i,i{+}1)$ (Fig.~\ref{triangle1}). We take the quadruple cut $\langle ABi{-}2i{-}1\rangle=\langle ABi{-}1i\rangle=\langle ABii{+}1\rangle=\langle ABI_\infty\rangle=0$, which corresponds to that of two-mass-hard box with a dual momentum sent to $I_{\infty}$. There are two solutions $(AB)_1=(i{-}1 I_{\infty} \cap \bar{i})$ and $(AB)_2=(i I_{\infty} \cap \overline{(i{-}1)})$, which give two intersection points on the line $I_{\infty}$: $I_{\infty} \cap \bar i$ and $I_{\infty} \cap \overline{(i{-}1)}$. Note that $\cap$ denotes the intersection of lines and/or planes. In this case, we intersect the line $I_\infty$ with, {\it e.g.}, the plane $\bar{i}:=(i{-}1, i, i{+}1)$. Such intersection points will be useful later, though for this case there are not enough points to build cross-ratios from (note the absence of dimensionless ratios of kinematic invariants).

For the two-mass triangle $I_3(i,i{+}1,j)$ (Fig.~\ref{triangle2}), the quadruple cut is equivalent to that of a three-mass box with $I_\infty$, and we find two intersection points on $I_\infty$ (from two solutions of the Schubert problem): $I_\infty\cap\bar{i}$ and $I_\infty\cap(ij{-}1 j)$. On the other hand, on the {\it internal line} $(AB)_1=(I_\infty\cap\bar i,(j{-}1j)\cap\bar i)$, there are four points from intersecting it with four lines. For the first time, we have a configuration with non-trivial cross-ratios: it can be represented by the Grassmannian $G(2,4)$ modulo torus action (or equivalently, the rank-one cluster algebra $A_1$), and the $2\times 4$ matrix reads
\begin{equation}
\left(\begin{array}{cccc}
     1  & \langle i{-}1ij{-}1j\rangle &  \langle ii{+}1j{-}1j\rangle & 0\\
     0  & {-}\langle i{-}1iI_\infty\rangle & -\langle ii{+}1I_\infty\rangle & 1
\end{array}
\right),\end{equation}
where we have chosen the gauge fixing by expressing any point on $I_\infty$ as a linear combination $I_{\infty} \cap \bar{i}$ and $(j{-}1 j) \cap \bar{i}$. We can build cross-ratios using the four columns:
\begin{equation}
    \mathcal U:=\frac{(12)(34)}{(13)(24)}=\frac{\langle i{-}1 i I_{\infty}\rangle \langle i i{+}1 j{-}1 j\rangle}{\langle i{-}1 i j{-}1 j\rangle \langle i i{+}1 I_{\infty} \rangle}=\frac{m_1^2}{m_2^2} 
\end{equation}
and $\mathcal V:=\frac{(14)(23)}{(13)(24)}=1-\mathcal U=1-m_1^2/m_2^2$. This $A_1$ cluster algebra is nothing but the alphabet of two-mass triangle.

Similarly, the three-mass triangle can be obtained from the four-mass box with $(l l{-}1)\to I_{\infty}$ (Fig.~\ref{triangle3}), and in addition to points on external lines, we get four points on each of the two cut-solutions (internal lines) $(AB)_{1,2}$. The cross-ratios of the four points on each internal line give the well-known alphabet $\{z, 1-z, \bar{z}, 1-\bar{z}\}$ of two $A_1$'s (details can be found in appendix VII). Next we show that these elements from triangles can be combined to give cross-ratios for general one-loop integrals. 

\subsection{Symbol letters of box integrals from geometries}

\begin{figure}[htbp]
    \centering
    \subcaptionbox{One-mass box\label{box1mint}}[0.4\linewidth][c]{
        \begin{tikzpicture}[scale=0.2,baseline={(current bounding box.center)}]
            \draw[black,thick] (0,0)--(4,0)--(4,4)--(0,4)--cycle;
            \draw[black,thick] (0,0)--(-1.41,-1.41);
            \draw[black,thick] (4,0)--(5.41,-1.41);
            \draw[black,thick] (4,4)--(5.41,5.41);
            \draw[black,thick] (-2,4)--(0,4)--(0,6);
            \filldraw[black] (5.41,5.41) node[anchor=south west] {{$2$}};
            \filldraw[black] (5.41,-1.41) node[anchor=north west] {{$3$}};
            \filldraw[black] (-2,4) node[anchor=east] {{$5$}};
            \filldraw[black] (0,6) node[anchor=south] {{$1$}};
            \filldraw[black] (-1.41,-1.41) node[anchor=north east] {{$4$}};
            \filldraw[black] (-4.24,-2.83) node[anchor=center] {{$P_4$}};
            \filldraw[black] (-2.83,6.83) node[anchor=center] {{$P_1$}};
            \filldraw[black] (8.24,6.83) node[anchor=center] {{$P_2$}};
            \filldraw[black] (8.24,-2.83) node[anchor=center] {{$P_3$}};
        \end{tikzpicture}
    }
    \subcaptionbox{Intersections on $I_\infty$\label{box1mpq}}[0.55\linewidth][c]{
        \begin{tikzpicture}[scale=0.2,baseline={([yshift=-0.7cm]current bounding box.south)}]
            \draw[black,ultra thick] (0,0) -- (21,0);
            \filldraw[black] (10.5,2) node[anchor=center] {{$I_\infty$}};
            \node[fill=red,circle,inner sep=1.5] at (2.5,0) {};
            \filldraw[red] (4,0) node[anchor=north east] {{$Z_\text{2-loop}$}};
            \node[fill=black,circle,inner sep=1.5] at (5.7,0) {};
            \filldraw[black] (5.7,0) node[anchor=north] {{$[345]$}};
            \node[fill=black,circle,inner sep=1.5] at (8.9,0) {};
            \filldraw[black] (8.9,0) node[anchor=north] {{$[245]$}};
            \node[fill=blue,circle,inner sep=1.5] at (12.1,0) {};
            \filldraw[blue] (12.1,0) node[anchor=north] {{$[234]$}};
            \node[fill=black,circle,inner sep=1.5] at (15.3,0) {};
            \filldraw[black] (15.3,0) node[anchor=north] {{$[124]$}};
            \node[fill=black,circle,inner sep=1.5] at (18.5,0) {};
            \filldraw[black] (18.5,0) node[anchor=north] {{$[123]$}};
        \end{tikzpicture}
    }

    \subcaptionbox{Two-mass-easy box\label{box2meint}}[0.4\linewidth][c]{
        \begin{tikzpicture}[scale=0.2,baseline={(current bounding box.center)}]
            \draw[black,thick] (0,0)--(4,0)--(4,4)--(0,4)--cycle;
            \draw[black,thick] (0,0)--(-1.41,-1.41);
            \draw[black,thick] (4,-2)--(4,0)--(6,0);
            \draw[black,thick] (4,4)--(5.41,5.41);
            \draw[black,thick] (-2,4)--(0,4)--(0,6);
            \filldraw[black] (-2,4) node[anchor=east] {{$6$}};
            \filldraw[black] (0,6) node[anchor=south] {{$1$}};
            \filldraw[black] (5.41,5.41) node[anchor=south west] {{$2$}};
            \filldraw[black] (6,0) node[anchor=west] {{$3$}};
            \filldraw[black] (4,-2) node[anchor=north] {{$4$}};
            \filldraw[black] (-1.41,-1.41) node[anchor=north east] {{$5$}};
            \filldraw[black] (-4.24,-2.83) node[anchor=center] {{$P_4$}};
            \filldraw[black] (-2.83,6.83) node[anchor=center] {{$P_1$}};
            \filldraw[black] (8.24,6.83) node[anchor=center] {{$P_2$}};
            \filldraw[black] (6.83,-2.83) node[anchor=center] {{$P_3$}};
        \end{tikzpicture}
    }
    \subcaptionbox{Intersections on $I_\infty$\label{box2mepq}}[0.55\linewidth][c]{
        \begin{tikzpicture}[scale=0.2,baseline={([yshift=-0.7cm]current bounding box.south)}]
            \draw[black,ultra thick] (0,0) -- (24.5,0);
            \filldraw[black] (12.25,2) node[anchor=center] {{$I_\infty$}};
            \node[fill=red,circle,inner sep=1.5] at (1,0) {};
            \filldraw[red] (1,0) node[anchor=north] {{$\epsilon_-$}};
            \node[fill=black,circle,inner sep=1.5] at (4.2,0) {};
            \filldraw[black] (4.2,0) node[anchor=north] {{$[456]$}};
            \node[fill=black,circle,inner sep=1.5] at (7.4,0) {};
            \filldraw[black] (7.4,0) node[anchor=north] {{$[256]$}};
            \node[fill=black,circle,inner sep=1.5] at (10.6,0) {};
            \filldraw[black] (10.6,0) node[anchor=north] {{$[245]$}};
            \node[fill=black,circle,inner sep=1.5] at (13.8,0) {};
            \filldraw[black] (13.8,0) node[anchor=north] {{$[235]$}};
            \node[fill=black,circle,inner sep=1.5] at (17,0) {};
            \filldraw[black] (17,0) node[anchor=north] {{$[125]$}};
            \node[fill=black,circle,inner sep=1.5] at (20.2,0) {};
            \filldraw[black] (20.2,0) node[anchor=north] {{$[123]$}};
            \node[fill=red,circle,inner sep=1.5] at (23.4,0) {};
            \filldraw[red] (23.4,0) node[anchor=north] {{$\epsilon_+$}};
        \end{tikzpicture}
    }
    \caption{Some intersection points for integrals with one-mass and two-mass-easy box kinematics}
    \label{boxes}
\end{figure}
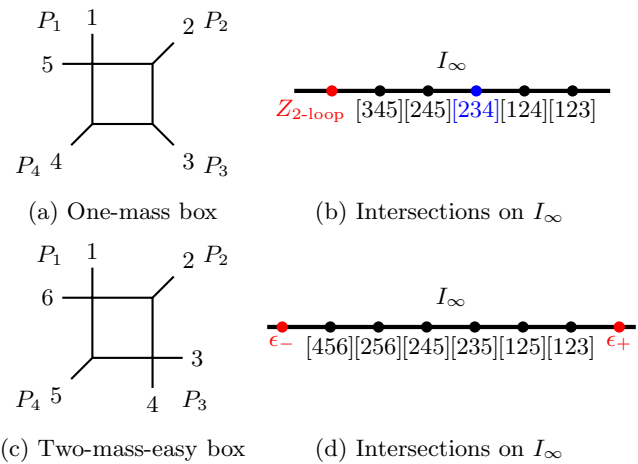

We first show that the symbol letters of any one-loop box integral can be obtained as cross-ratios. For the one-mass box with $m_1^2\neq 0$ (Fig.~\ref{box1mint}), in twistor space we can cut $\langle AB 12\rangle$, $\langle AB 23\rangle$, $\langle AB 34\rangle$, $\langle AB 45\rangle$, and $\langle AB I_{\infty}\rangle$~\footnote{The zero-mass box is essentially trivial: the four points on $I_{\infty}$ give $A_1$ cross-ratios $(\mathcal U,\mathcal V)=(s/(s+t), t/(s+t))$, as we review in appendix VII.A.}. It suffices to consider cross-ratios constructed from pairs of triangles: by cutting $\langle ABI_\infty\rangle$ and three other inverse propagators at a time, we obtain two intersection points on $I_\infty$ for each triangle sub-topology, so that a pair of triangles gives an $A_1$. Equivalently, the four triangle sub-topologies produce $5$ points on $I_{\infty}$ altogether, which are the intersections of $I_\infty$ with the planes $(123), (124), (234), (245)$ and $(345)$ as shown in Fig.~\ref{box1mpq}, where $[ijk]:=I_\infty\cap(ijk)$~\footnote{We comment that for the one-mass and the two-mass-easy box, if we take $I_{\infty}=(n{+}1, n{+}2)$ for $n=5,6$ and require all $n{+}2$ twistors to be in $G_+(4,n{+}2)$, then the intersection points are ordered as shown in Fig.~\ref{boxes} and the $A_2, A_3$ alphabets become positive.}. The independent cross-ratios of these $5$ points can be chosen to be the $u$-variables of the $A_2$ cluster algebra satisfying the $u$-equations~\cite{Arkani-Hamed:2019plo} (see appendix VII for details):
\begin{align}
&1-u_{1,3}=\frac{\langle 23 45\rangle\langle 12 I_{\infty}\rangle}{\langle 12 45\rangle\langle 23 I_{\infty}\rangle}=\frac{s}{m_1^2},\nonumber\\
&1-u_{2,4}=\frac{\langle 1234 \rangle\langle 45 I_{\infty}\rangle}{\langle 12 45\rangle\langle 34 I_{\infty}\rangle}=\frac{t}{m_1^2}, {\it etc.}
\end{align}
The upshot is that the $5$ multiplicatively independent cross-ratios are given by ratios of $6$ letters $\{m_1^2, s, t, s-m_1^2, t-m_1^2, s+t-m_1^2\}$. This $A_2$ alphabet is well known from canonical differential equations (CDE) calculations~\cite{Henn:2013pwa,Henn:2014qga,Chicherin:2020umh}, and it is satisfying that we get it directly from the $5$ points on $I_{\infty}$ by considering all possible quadruple cuts!

Similarly for higher-mass boxes, we compute cross-ratios of intersections by considering all possible quadruple cuts. Beautifully, for two-, three-, and four-mass boxes, we find 6, 7, and 8 intersections on $I_{\infty}$, which lead to cross-ratios forming $A_3$, $A_4$, and $A_5$, respectively. For example, for the two-mass-easy box (Fig.~\ref{box2meint}), the 6 intersections on $I_{\infty}$ as shown in Fig.~\ref{box2mepq}, giving 9 cross-ratios which are ratios of $m_1^2, m_3^2, s, t$ and
\begin{equation*}
s{-}m_1^2, t{-}m_1^2, s{-}m_3^2, t{-}m_3^2, m_1^2+m_3^2{-}s{-}t, st{-}m_1^2m_3^2. 
\end{equation*}
Beyond two-mass-easy case, some cross-ratios from internal lines $(AB)$ with (at least) four intersections become independent of those from $I_{\infty}$, and we need to include them as well. For the two-mass-hard box, one additional line with four points is needed, and the alphabet is given by $A_3 \cup A_1$ with $11$ letters. For the three-mass box, the alphabet can be generated as $A_4 \cup (A_1)^3$ containing $18$ letters (we give details in appendix VII.A).

The above results for up to three-mass box cases agree with~\cite{Chicherin:2020umh}; for four-mass box case, the complete two-loop (double-box) alphabet, which includes one-loop alphabet, has been recently computed and reproduced from twistor geometries of DCI integrals~\cite{He:2022ctv}. We emphasize that in every case, the space spanned by cross-ratios of our geometric construction exactly agree with that spanned by letters of CDE for corresponding master integrals (with uniform transcendental weights) for the family; we review such CDE and alphabets in appendix IX. 

\subsection{Symbol letters of integrals beyond boxes}

Now, we move on to one-loop integrals which depend on five or more dual momenta, and it suffices to consider pentagons since they saturate the basis of one-loop integrals in $D=4-2\epsilon$ \cite{Bern:1993kr}. A new feature is the appearance of parity-odd letters of the form $\frac{a+b\mathop{\rm tr}_5}{a-b\mathop{\rm tr}_5}$, where $\mathop{\rm tr}_5=4i\epsilon_{\mu\nu\rho\sigma}P_1^\mu P_2^\nu P_3^\rho P_4^\sigma=\sqrt{\det(2P_i\cdot P_j)_{i,j=1,\cdots,4}}$ changes sign under a space-time parity transformation. We remark that $\mathop{\rm tr}_5$, an irreducible square root of Mandelstam variables, is rationalized by momentum twistors.

The procedure goes exactly as before: we take all possible quadruple cuts, including $\langle AB I_\infty\rangle$, and compute cross-ratios on $I_\infty$ or on cut-solutions $(AB)$ with at least four intersection points. A cross-ratio has the generic form $a+b\mathop{\rm tr}_5$ with $a,b$ independent of $\mathop{\rm tr}_5$~\footnote{Whenever one sees a denominator $c+d\mathop{\rm tr}_5$, just multiply both the numerator and the denominator with $c-d\mathop{\rm tr}_5$ and use $\mathop{\rm tr}_5^2=\det(2P_i\cdot P_j)$.}. However, not all of them are allowed: the definition of momentum twistors is chiral and it is possible that the cross-ratios we find span a space which includes $a+b \mathop{\rm tr}_5$ but not $a-b \mathop{\rm tr}_5$. Since the complete symbol alphabet of a family of integrals (which only depend on the set of propagators) should be parity-invariant, the letters must be either parity-even (independent of $\mathop{\rm tr}_5$) or parity-odd (of the form $\frac{a+b\mathop{\rm tr}_5}{a-b\mathop{\rm tr}_5}$). 
Therefore, we have to select the parity-invariant subspace of the span of all possible cross-ratios, a complication which is non-existent for boxes. In this way we will produce all parity-odd letters, but not $\mathop{\rm tr}_5$ itself since it is the LS of a $D=6$ pentagon. However, the letter $\mathop{\rm tr}_5$ also drops out nicely in known physical quantities.

Consider the zero-mass pentagon first, which depends on $\{s_{i,i+1}\}_{i=1,\cdots,5}$ and has five one-mass box sub-topologies. The above procedure yields 15 parity-even letters (more precisely, 14 dimensionless ratios of them) and 5 parity-odd ones, exactly matching the one-loop result in \cite{Chicherin:2017dob} (except for $\mathop{\rm tr}_5$ itself)
\begin{equation}
    \begin{split}
        \Big\{& \frac{s_{12}s_{23}-s_{23}s_{34}+s_{34}s_{45}-s_{45}s_{51}-s_{51}s_{12}+\mathop{\rm tr}_5}{s_{12}s_{23}-s_{23}s_{34}+s_{34}s_{45}-s_{45}s_{51}-s_{51}s_{12}-\mathop{\rm tr}_5},\\
        & s_{12},\ s_{12}-s_{45},\ s_{12}+s_{23}-s_{45},\ \text{+ cyclic}\Big\}.
    \end{split}
\end{equation}
Interestingly, the union of five $A_2$ alphabets from sub-topologies already covers all parity-even letters, and the parity-odd letters arise from mixing two triangles from different boxes. Moreover, as a consistency check, the product of numerator and denominator of any parity-odd letter belongs to the parity-even sector of the alphabet.

It is straightforward to generalize to pentagons with massive corners. In the one-mass case, we find exactly 23 parity-even letters ({\it i.e.}, 22 ratios) and 6 parity-odd ones, matching \cite{Abreu:2020jxa} except for $\mathop{\rm tr}_5$ itself. Again, all parity-even letters are spanned by the union of box sub-topologies, and parity-odd letters arise from mixing triangles from different boxes.

For more massive pentagons, we present analysis for two-mass-easy and two-mass-hard cases in the appendix VII.B. For completeness, we also record their one-loop CDE in an ancilarly file, which to our knowledge has not been explicitly presented in the literature. In all these cases, cross-ratios from geometric configurations successfully cover the alphabets from CDE.

\section{Two-loop geometries and symbology}
Next we exploit twistor geometries associated with two-loop integrals to predict their symbol letters: 
the basic strategy is to consider two-loop LSs which usually generate new intersection points, thus producing more cross-ratios, on top of the one-loop intersections. Such cross-ratios naturally include as factors two-loop LS themselves, but more importantly they produce new letters involving them, {\it e.g.}, algebraic letters when the LS is a square root. We outline our results and provide more details in the appendix. 
\begin{figure}[htbp]
\begin{subfigure}{0.3\linewidth}
 \begin{tikzpicture}[scale=0.15]
                \draw[black,thick] (0,5)--(-5,5)--(-5,0)--(0,0)--cycle;
                \draw[black,thick] (0,5)--(1.4374,6.2624);
                  \draw[black,thick] (-4.9367,0.0496)--(0.0518,5.0305);
                \draw[black,thick] (2.1042,-1.5613)--(0.0733,-0.1445);
                \draw[black,thick] (-6.93,5.52)--(-5,5)--(-5.52,6.93);
                \draw[black,thick] (-5,0)--(-6.6028,-1.6224);
                \filldraw[black] (0.7164,5.8642) node[anchor=south west] {{$2$}};
                \filldraw[black] (2.1567,-1.5161) node[anchor=north] {{$3$}};
                \filldraw[black] (-6.93,6) node[anchor=east] {{$5$}};
                \filldraw[black] (-5.52,6.93) node[anchor=south] {{$1$}};
                \filldraw[black] (-6.2212,-1.1446) node[anchor=north east] {{$4$}};
                   \filldraw[black] (-7.3471,-0.8091) node[anchor=north east] {{$P_4$}};
                     \filldraw[black] (-5.7402,10.8914) node[anchor=north east] {{$P_1$}};
                       \filldraw[black] (8.0314,9.7102) node[anchor=north east] {{$P_2$}};
                         \filldraw[black] (7.8753,-0.2624) node[anchor=north east] {{$P_3$}};
            \end{tikzpicture}
            \caption{}\label{slashboxesa}
\end{subfigure}
\begin{subfigure}{0.3\linewidth}
 \begin{tikzpicture}[scale=0.15]
                \draw[black,thick] (0,5)--(-5,5)--(-5,0)--(0,0)--cycle;
                \draw[black,thick] (0,5)--(1.4374,6.2624);
                  \draw[black,thick] (-4.9367,0.0496)--(0.0518,5.0305);
                \draw[black,thick] (1.2225,-1.9266)--(-0.0344,-0.1974)--(2.4069,-0.4638);
                \draw[black,thick] (-6.93,5.52)--(-5,5)--(-5.52,6.93);
                \draw[black,thick] (-5,0)--(-6.6028,-1.6224);
                \filldraw[black] (0.7164,5.8642) node[anchor=south west] {{$2$}};
                \filldraw[black] (1.2505,-1.8954) node[anchor=north] {{$4$}};
                                \filldraw[black] (2.463,-0.5512) node[anchor=west] {{$3$}};
                \filldraw[black] (-6.93,6) node[anchor=east] {{$6$}};
                \filldraw[black] (-5.52,6.93) node[anchor=south] {{$1$}};
                \filldraw[black] (-6.2212,-1.1446) node[anchor=north east] {{$5$}};
                   \filldraw[black] (-7.3471,-0.8091) node[anchor=north east] {{$P_4$}};
                     \filldraw[black] (-5.7402,10.8914) node[anchor=north east] {{$P_1$}};
                       \filldraw[black] (8.0314,9.7102) node[anchor=north east] {{$P_2$}};
                         \filldraw[black] (6.9936,-0.6277) node[anchor=north east] {{$P_3$}};
            \end{tikzpicture}
            \caption{}\label{slashboxesb}
\end{subfigure}
                  \begin{subfigure}{0.3\linewidth}
                   \begin{tikzpicture}[scale=0.15]
                \draw[black,thick] (0,5)--(-5,5)--(-5,0)--(0,0)--cycle;
                \draw[black,thick] (0,5)--(1.4374,6.2624);
                                \draw[black,thick] (-2.5,2.5)--(-4.9893,4.9737);
                  \draw[black,thick] (-4.9367,0.0496)--(-0.033,4.9966);
                \draw[black,thick] (1.2225,-1.9266)--(-0.1293,0.0321)--(2.4069,-0.4638);
                \draw[black,thick] (-6.93,5.52)--(-5,5)--(-5.52,6.93);
                \draw[black,thick] (-5,0)--(-6.6028,-1.6224);
                \filldraw[black] (0.7164,5.8642) node[anchor=south west] {{$2$}};
                \filldraw[black] (1.2505,-1.8954) node[anchor=north] {{$4$}};
                                \filldraw[black] (2.463,-0.5512) node[anchor=west] {{$3$}};
                \filldraw[black] (-6.93,6) node[anchor=east] {{$6$}};
                \filldraw[black] (-5.52,6.93) node[anchor=south] {{$1$}};
                \filldraw[black] (-6.2212,-1.1446) node[anchor=north east] {{$5$}};
                   \filldraw[black] (-7.3471,-0.8091) node[anchor=north east] {{$P_4$}};
                     \filldraw[black] (-5.7402,10.8914) node[anchor=north east] {{$P_1$}};
                       \filldraw[black] (8.0314,9.7102) node[anchor=north east] {{$P_2$}};
                         \filldraw[black] (6.9936,-0.6277) node[anchor=north east] {{$P_3$}};
            \end{tikzpicture}
            \caption{}
            \label{slashboxesc}
                  \end{subfigure}          
            \caption{One-mass \subref{slashboxesa} and two-mass-easy \subref{slashboxesb} slashed- boxes, and a three-loop wheel integral \subref{slashboxesc}}
             \label{slashboxes}
\end{figure}
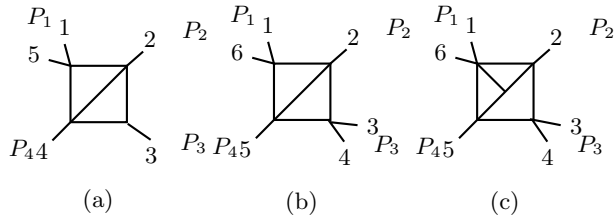

Let us first consider the zero-mass pentagon kinematics, where we have already found $20$ even letters and $5$ odd ones from points on $I_{\infty}$ generated by one-loop integrals. At two loops, the simplest non-trivial topology is a double-triangle (or slashed-box, Fig.~\ref{slashboxesa}), which belongs to the sub-topology of one-mass box kinematics. We need the LS configuration where $(AB)$ intersect with $(CD)$, $(AB)$ with $(12)$, $(45)$ and $I_{\infty}$, as well as $(CD)$ with $(23)$, $(34)$ and $I_{\infty}$. Writing $I_{\infty}=(PQ)$ with two arbitrary reference points, it is easy to compute the LS by parametrizing $Z_A=\alpha_1 Z_1+\beta_1 Z_P+Z_Q$, $Z_B=\gamma_1 Z_1+\delta_1 Z_P+Z_2$ and $Z_C=\alpha_2 Z_2+\beta_2 Z_P+Z_Q$, $Z_D=\gamma_2 Z_2+\delta_2 Z_P+Z_3$. After taking residue in all $8$ variables, we find two intersections on $I_\infty$, corresponding to solutions where $\beta_1=\beta_2=-\frac{\langle Q234\rangle}{\langle P234\rangle}$ or $\beta_1=\beta_2=\frac{\langle Q(12)(3P)(45)\rangle}{\langle P(12)(3Q)(45)\rangle}$. Note that $\beta_1=\beta_2$ means that three lines $(AB)$, $(CD)$ and $I_{\infty}$ intersect at the same point. One of the intersections is a familiar one-loop intersection, $I_{\infty} \cap \bar{3}$ (blue in Fig.~\ref{box1mpq}), and the other is new (red in Fig.~\ref{box1mpq}), which reads
\[Z_{\rm 2-loop} \propto\langle Q(12)(3P)(45)\rangle Z_P+\langle P(12)(3Q)(45)\rangle Z_Q.\] The new point thus introduces new cross-ratios, {\it e.g.}, any cross-ratio containing the minor of these two points is proportional to $s+t$, which is the (inverse of) LS of this integral. This is a general phenomenon: for any two-loop integral, the minor of two of the intersections on $I_{\infty}$, which are generated by the maximal cut, is always proportional to the inverse of its LS, and generally should be considered as a new letter~\footnote{In the special cases where the two-loop intersection points coincide with one-loop intersections, the two-loop LS does not give any new letter. For example, this is the case for the triangle-box with the massive corner $\{5,1\}$ on the triangle side.}.

By going through all possible two-loop integrals with one-mass box kinematics, we find that the only new intersection is $Z_{\rm 2-loop}$ above, and the cross-ratios of the $6$ points (together with $5$ from one-loop) only contains one new letter, which is $s+t$~\footnote{Note that the alphabet of these $6$ points is not $A_3$ since they only depend on $2$ variables, see appendix VII.B for more details.}. For the zero-mass pentagon kinematics, we have $5$ such sub-topologies, resulting in $5$ new even letters, $\{s_{12}+ s_{23}, s_{23}+s_{34}, \cdots, s_{51}+s_{12}\}$. 

To obtain the parity-odd letters, we need to consider integrals depending on all $5$ dual momenta, such as triangle-boxes or triangle-pentagons. Nicely we find no new intersections on $I_\infty$, thus we conclude that, up to two loops, the geometries give $20+5$ symbol letters, which agrees perfectly with CDE results~\cite{Chicherin:2017dob}, if we include $W_{31}:={\rm tr}_5$ itself. 

Next we move to the one-mass pentagon and consider its sub-topologies, where the first non-trivial two-loop case is the two-mass-easy double-triangle (Fig.~\ref{slashboxesb}). Completely analogously to the previous case, we find two new LS intersections on $I_\infty$ (red in Fig.~\ref{box2mepq}) whose minor gives the two-loop square root $\Delta_{nc}=\sqrt{(s+t)^2-4 m_1^2 m_3^2}$ proportional to the (inverse of) LS~\cite{Abreu:2020jxa}. Moreover, their cross-ratios with one-loop intersections produce more algebraic letters. In total we have $6+2=8$ points, and their cross-ratios cover a $3$-dim subspace of algebraic letters involving $\Delta_{nc}$, which are spanned by
{\small\begin{equation}\label{L3}
L_1{=}\frac{s+t+\Delta_{nc}}{s+t-\Delta_{nc}},L_2{=}\frac{s-t+\Delta_{nc}}{s-t-\Delta_{nc}},L_3{=}\frac{-2m_1^2+s+t+\Delta_{nc}}{-2m_1^2+s+t-\Delta_{nc}}.
\end{equation}}%
These $9+3$ letters (as well as the irrelevant letter $\Delta_{nc}$) are exactly those computed from CDE. Again the product of numerator and denominator of $L_i$ gives (rational) letters in $A_3$. Similarly for two-mass-hard box sub-topologies, we find $4$ new rational LS letters from double-triangle and triangle-box integrals, which together with the $11$ one-loop letters give the full two-loop two-mass-hard box alphabet of $15$ letters. We have also reproduced all two-loop letters for three-mass and four-mass box kinematics~\cite{Dlapa:2021qsl,He:2022ctv}, and we record the construction of two-loop letters up to three-mass box kinematics from twistor geometries in appendix VIII.B.

We reproduce the full two-loop alphabet for the one-mass pentagon kinematics~\cite{Abreu:2020jxa}: by taking the union of five box sub-topologies, only $6$ relevant letters and $9$ irrelevant letters are missing, which can be generated by a few genuinely $5$-point integrals, as explained in detail in appendix VIII.A. Moreover, we have generated a preliminary two-loop alphabet for pentagon kinematics with two masses~\footnote{It is easy to see that some two-loop integrals with three-mass pentagon kinematics, such as box-triangle integral with two massless corners in the middle, evaluate to functions beyond MPLs. }, in addition to one-loop alphabets recorded in the ancillary file. However, more works are needed to select the correct letters, and we leave this and comparison with CDE to future works. 

\section{The origin and applications of (extended) cluster algebras} 

The symbol letters of certain integrals in $D=4-2 \epsilon$ are organized into cluster algebras and extensions (c.f.~\cite{Chicherin:2020umh}), and similarly for numerous examples of multi-loop finite integrals~\cite{He:2021esx,*He:2021fwf,*He:2021eec} (both DCI and non-DCI ones with $I_{\infty}$). What we have found here provides an explanation for these cluster algebraic structures as well as some clarifications. Our proposal makes it clear that the alphabet of any integral may be understood as the union of various $A_1$ cluster algebras ($4$ points on a line). Clearly, $m$ independent points on a line leads to $A_{m{-}3}$; {\it e.g.}, alphabets of one-loop one-mass and two-mass-easy box integrals correspond to $m=5,6$, respectively. 

As we have seen for one-loop boxes and pentagons, cross-ratios on more than one line are needed, and the union of such type-$A$ cluster algebras usually give subsets or certain limits of other finite-type cluster algebras. For the two-mass-hard or three-mass box, our $A_3\cup A_1$ and $A_4 \cup A_1^3$ are subsets of the $C_3$ and $C_4$ in~\cite{Chicherin:2020umh}, respectively. Beyond one loop, it becomes clear that extensions of cluster algebras are needed: {\it e.g.}, for the two-mass-easy box kinematics, we find two new points on $I_{\infty}$ at two-loop which nicely give $3$ additional algebraic letters, and we have seen similar phenomena with integrals with higher points and more massive corners.

Though these alphabets require extensions of cluster algebras ({\it e.g.}, with algebraic letters), they all eventually come from unions of $A_1$'s, just like those of DCI integrals~\cite{Yang:2022gko}.
To illustrate the power of twistor geometries and extended cluster algebras, we apply them to two- and three-loop integrals with two-mass-easy box kinematics (Fig.~\ref{slashboxesb},~\ref{slashboxesc}), which follow from DCI two-mass-easy pentagon kinematics (known as $D_3=A_3$~\cite{He:2021eec}) by sending a point to $I_{\infty}$. Compared to the two-loop double-triangle, the LS of three-loop ``wheel" integral does not produce any new intersection, thus the latter must still have an alphabet of the extended $A_3$ cluster algebra with $9$ rational letters and the $3$ algebraic ones in \eqref{L3}. Remarkably, we find that both two- and three-loop integrals satisfy nice differential equations in $D=4$:
\begin{equation}\label{DE23loops}
{\rm d}I^{(\ell)}=\sum_{i=1}^3F^{(2\ell{-}1)}_i(A_3)\,{\rm d}\log L_i, \quad {\text for}~\ell=2,3 
\end{equation}
where the two integrals are denoted as $I^{(\ell)}$ with $\ell=2,3$, and very nicely here $F^{(2\ell{-}1)}_{i=1,2,3}$ are weight-$(2\ell{-}1)$ functions of the $A_3$ alphabet~\footnote{Note that these integrals can be obtained as certain limits of two-loop $n=10$ and three-loop $n=9$ integrals, respectively, which evaluate to functions beyond MPL; for $\ell=2$ the $\log L_i$'s follow from the limit of ``last entries'' of the $n=10$ double-box, which are elliptic integrals~\cite{Kristensson:2021ani}.}. 

The $\ell=2$ case of \eqref{DE23loops} can be derived from CDE for the integral truncated at ${\cal O}(\epsilon)$ and $F_i^{(3)}$ are given by master integrals in the sector at leading order~\cite{Henn:2014lfa}. While we do not have the CDE for $I^{(3)}$, it has been obtained from the bootstrapped DCI results in~\cite{He:2021eec}, which shows that this rather constraining alphabet is valid at least for this sector at three loops! Similar differential equations have been found for three-mass and four-mass two-loop integrals~\cite{Dlapa:2021qsl,He:2022ctv}, where we have $4$ and $5$ algebraic letters as last entries in $D=4$, respectively.

\section{Generalizations to non-planar integrals}
Our method can be applied to integrals with non-planar kinematics. For example, if we consider the kinematics with one massive leg and $3$ massless legs without any particular ordering (which is also the kinematics for the three-point form factor~\cite{Dixon:2020bbt,*Dixon:2022rse}), the well-known $C_2$ alphabet (for integrals that evaluate to MPL) also comes from the union of $A$'s. Recall that for the one-mass box with lines $\{(12), (23), (34), (45)\}$, the alphabet is $A_2$ with $5$ points on $I_{\infty}$. Note that the kinematics stay the same for any ordering, {\it e.g.}, with lines $\{(13), (32), (24), (45)\}$, but we get a different $A_2$ for such a ordering. Most letters of these two $A_2$'s coincide, and the union gives $C_2$ with one more independent cross-ratio, which is in fact equivalent to the union of all $3!=6$ $A_2$'s. Similarly, by going through the different orderings of zero-mass pentagons (up to two loops), we find the $30$ letters of the non-planar alphabet~\cite{Chicherin:2018mue} (except for ${\rm tr}_5$). 

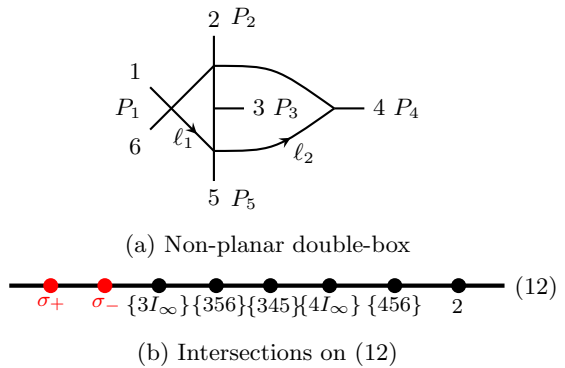
\begin{figure}[htbp]
    \centering
    \begin{subfigure}{0.8\linewidth}
        \centering
        \begin{tikzpicture}[scale=0.2,decoration={markings,mark=at position 0.6 with \arrow{stealth}}]
            \draw[black,thick] (0,4.83)--(0,-4.83);
            \draw[black,thick] (0,2.83)--(-4.24,-1.41);
            \draw[black,thick] (-4.24,1.41)--(-2.83,0);
            \draw[black,thick,postaction={decorate}] (-2.83,0)--(0,-2.83);
            \draw[black,thick] (0,0)--(2,0);
            \filldraw[black] (2,0) node[anchor=west] {{$3\ P_3$}};
            \draw[black,thick] (0,2.83) .. controls(4,2.83) .. (8,0);
            \draw[black,thick,postaction={decorate}] (0,-2.83) .. controls(4,-2.83) .. (8,0);
            \draw[black,thick] (8,0)--(10,0);
            \filldraw[black] (10,0) node[anchor=west] {{$4\ P_4$}};
            \filldraw[black] (0,4.83) node[anchor=south] {{$2$}};
            \filldraw[black] (0,-4.83) node[anchor=north] {{$5$}};
            \filldraw[black] (-4.24,-1.41) node[anchor=north east] {{$6$}};
            \filldraw[black] (-4.24,1.41) node[anchor=south east] {{$1$}};
            \filldraw[black] (2,4.84) node[anchor=south] {{$P_2$}};
            \filldraw[black] (2,-4.84) node[anchor=north] {{$P_5$}};
            \filldraw[black] (-4.24,0) node[anchor=east] {{$P_1$}};
            \filldraw[black] (-2,-2) node[anchor=center] {{$\ell_1$}};
            \filldraw[black] (6,-3) node[anchor=center] {{$\ell_2$}};
        \end{tikzpicture}
        \caption{Non-planar double-box}
        \label{nonpl1}
    \end{subfigure}
    
    \begin{subfigure}{0.8\linewidth}
        \begin{tikzpicture}[scale=1.2]
            \draw [black, ultra thick](0.6504,-5.2) -- (6.1371,-5.2);
            \node [fill=black,circle,inner sep=2pt] at (5.6286,-5.2) {};
            \node [fill=black,circle,inner sep=2pt] at (4.9213,-5.2) {};
            \node [fill=black,circle,inner sep=2pt] at (4.1989,-5.2) {};
            \node [fill=black,circle,inner sep=2pt] at (3.5438,-5.2) {};
            \node [fill=black,circle,inner sep=2pt] at (2.9435,-5.2) {};
            \node [fill=black,circle,inner sep=2pt] at (2.3096,-5.2) {};
            \node [fill=red,circle,inner sep=2pt] at (1.7096,-5.2) {};
            \node [fill=red,circle,inner sep=2pt] at (1.1096,-5.2) {};
            \node at (6.4894,-5.2289) {$(12)$};
            \node [black] at (5.6138,-5.4235) {\footnotesize $2$};
            \node [black] at (4.1979,-5.4201) {{\footnotesize $\{4I_\infty\}$}};
            \node [black] at (2.9821,-5.4237) {\footnotesize$\{356\}$};
            \node [black] at (2.3016,-5.423) {\footnotesize$\{3I_\infty\}$};
            \node [black] at (4.9189,-5.4186) {\footnotesize $\{456\}$};
            \node [black] at (3.5819,-5.431) {\footnotesize $\{345\}$};
            \node [red] at (1.1181,-5.398) {\footnotesize $\sigma_+$};
            \node [red] at (1.7296,-5.412) {\footnotesize $\sigma_-$};
        \end{tikzpicture}
        \caption{Intersections on $(12)$}
        \label{nonplb}
    \end{subfigure}
    \caption{A non-planar double-box integral with one-mass pentagon kinematics and the intersections on the line $(12)$}
    \label{fig:nonpl}
\end{figure}
For genuinely non-planar integrals, we propose an extension of our method which is inspired by
the study of dual conformal transformation beyond the planar limit~\cite{Bern:2017gdk,*Bern:2018oao,*Chicherin:2018wes}. Consider the non-planar double-box in Fig.~\ref{nonpl1}. By introducing $\ell_1=y_1-x_1$, $\ell_2=y_2-x_5$ and $P_i=x_{i, i{+}1}:=x_{i{+}1}-x_i$ for $i=1, \cdots, 5$ (it is a slight misuse of notation: here these $x_i$'s are not the same as those in the planar case), we have inverse propagators $(y_1-x_1)^2$, $(y_1-x_2)^2$, $(y_2-x_4)^2$, $(y_2-x_5)^2$, $(y_1-y_2)^2$ as well as $(y_1-y_2+x_{3,4})^2$. 
Since we are interested in the maximal cut, on the support of $(y_1-y_2)^2=0$, $(y_1-y_2+x_{3,4})^2=0$ is equivalent to 
$2 P_3\cdot (y_1-y_2)=(y_1-x_4)^2-(y_1-x_3)^2
-(y_1 \leftrightarrow y_2)=0$. 

We can rewrite these using momentum twistors: $y_1 \sim (AB)$, $y_2 \sim (CD)$, $x_1 \sim (56)$ and $x_i\sim (i{-}1i)$ for $i=2,\cdots,5$; equivalently we cut $\langle AB 12\rangle $, $\langle AB 56\rangle$, $\langle CD 34\rangle$, $\langle CD 45\rangle$, $\langle ABCD\rangle$, and a new factor 
$\langle AB I_{\infty} \rangle \langle CD~{\bar 3} \cap (3 I_{\infty}) \rangle - (AB \leftrightarrow CD)$.
Very nicely, we find that although the integrand has changed, the LS by cutting the above $6$ factors remains the same, which is proportional to the square root $\Sigma_5^{(1)}$, given in eq.(2.7) of~\cite{Abreu:2021smk}. 

Now we can consider intersection points on various external lines. For instance, on line $(12)$, the above two-loop non-planar LS produces $\{\sigma_+,\sigma_-\}$, and there are several intersections produced from one-loop Schubert problems (with one-mass pentagon kinematics) (Fig.~\ref{nonplb}, where we denote $\{ijk\}:=(12)\cap(ijk)$). By considering all possible cross-ratios $\frac{(\sigma_1,X_1)(\sigma_2,X_2)}{(\sigma_1,X_2)(\sigma_2,X_1)}$ with $X_1$ and $X_2$ one-loop intersections, we successfully recover the $5$ algebraic letters involving $\Sigma_5^{(1)}$ computed using CDE~\cite{Abreu:2021smk}. We plan to apply this method to the symbology of more general two-loop non-planar integrals, such as those in~\cite{Bourjaily:2019gqu}.

\section{Conclusions}
By computing geometric invariants associated with leading singularities in momentum twistor space, we provided a simple access to the symbol alphabets of general Feynman integrals. We have applied our method to state-of-art two-loop integrals with {\it e.g.} one-mass pentagon kinematics, and we expect to push the frontier to higher-mass five-point or even six-point processes, which are of great phenomenological interest. Based on known alphabets and powerful constraints on symbols such as the (extended) Steinmann relations \cite{Steinmann1960a,*Steinmann1960b, Caron-Huot:2016owq,He:2021mme}, we have seen illuminating examples of bootstrapping finite multi-loop integrals (c.f.~\cite{Henn:2018cdp,He:2021non,He:2021eec}), and it would be extremely interesting to extend this to $D=4-2\epsilon$. 

Our study explains the origin and limitations of cluster algebras in symbol alphabets, and it is important to explore these further especially for non-planar integrals. Another important direction is to pursue the ``symbology'' for integrals that evaluate to functions beyond MPL, where twistor geometries have been crucial in studying their {\it rigidity} and Calabi-Yau structures~\cite{Bourjaily:2018ycu,Bourjaily:2019hmc,Vergu:2020uur}. Loosely speaking, the ``letters" of elliptic MPL (which are elliptic integrals {\it etc.}) are expected to be geometric invariants associated with elliptic curves as well, which reduce to simpler ones (such as $\log L_i$ in \eqref{DE23loops}) in limits where they become MPL~\cite{Wilhelm:2022wow}. We leave the systematic study of such invariants for elliptic leading singularities~\cite{Bourjaily:2020hjv} to future works.  

Last but not least, we remark that our construction may be closely related to other methods for symbology, such as Landau analysis ({\it c.f.}~\cite{Landau:1959fi,*Dennen:2015bet,*Dennen:2016mdk,*Mizera:2021icv}) and the diagrammatic coproduct~\cite{Abreu:2017enx,*Abreu:2017mtm,*Abreu:2019wzk,*Abreu:2021vhb}. 
It would be highly desirable to find precise relations among these methods. 

We thank Nima Arkani-Hamed, Zhenjie Li, Chi Zhang and Yang Zhang for inspiring discussions and collaborations on related projects. This research is supported in part by National Natural Science Foundation of China under Grant No. 11935013,11947301,12047502,12047503.

\bibliography{main}

\widetext

\begin{center}
\textbf{\large Supplemental materials}
\end{center}

\section{Review and details of one-loop geometries and alphabets}

\begin{figure}[H]
    \centering
    \begin{tikzpicture}[scale=0.55]
        \draw[black,ultra thick](-0.4788,4.1896)--(5.1391,1.0089);
        \draw[black,ultra thick](3.8192,0.6953)--(4.188,5.4454);
        \draw[black,ultra thick](-3.7979,-0.3736)--(-2.8629,5.6045);
        \draw[black,ultra thick](-4.4282,5.2568)--(2.2296,1.5794);
        \draw[blue,thick](-5.6093,5.5009)--(6.9982,0.4829);
        \draw[blue,thick](-6.6967,-0.6177)--(6.2357,5.7135);
        \node [fill=blue,circle,inner sep=2pt] at (1.2283,3.2619) {};
        \node [fill=blue,circle,inner sep=2pt] at (-3.5686,0.9021) {};
        \node [fill=blue,circle,inner sep=2pt] at (4.117,4.6508) {};
        \node [fill=blue,circle,inner sep=2pt] at (0.109,2.7062) {};
        \node [fill=blue,circle,inner sep=2pt] at (-3.0172,4.4971) {};
        \node [fill=blue,circle,inner sep=2pt] at (3.8906,1.7142) {};
        \node at (-3.5,5.2) {2};
        \node at (-3.3955,-0.0915) {1};
        \node at (2.1872,1.1686) {3};
        \node at (-0.6773,4.4194) {4};
        \node at (4.1172,1.2837) {5};
        \node at (4.4083,5.5208) {6};
        \node [blue] at (-6.4662,5.1821) {$(AB)_1=(25)$};
        \node [blue] at (6.1991,6.0584) {$(AB)_2=(\bar2\cap\bar5)$};
        \node [blue] at (1.2332,3.8518) {$(45)\cap\bar2$};
        \node [blue] at (-5,1.1429) {$(12)\cap\bar5$};
        \node [blue] at (5.0556,4.2744) {$(56)\cap\bar2$};
        \node [blue] at (0.0396,1.9) {$(23)\cap\bar5$};
    \end{tikzpicture}
    \caption{Quadruple cut of the DCI integral $I_4(2,3,5,6)$.}
    \label{dci2me}
\end{figure}
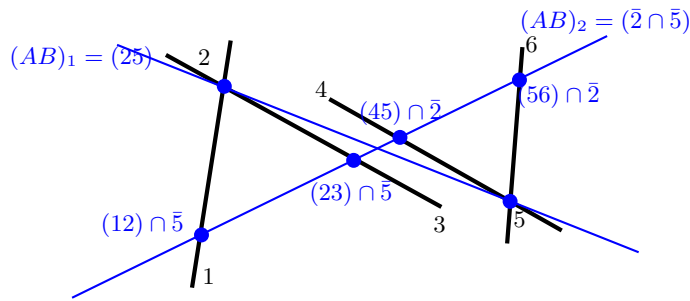

Let us start by highlighting some basic facts of Schubert problems (see~\cite{Arkani-Hamed:2010pyv} for a more detailed review) and their geometric invariants. Recall that a quadruple cut leads to a Schubert problem of intersecting the internal (loop) line $(AB)$ with four external lines, or dual momenta. For example, consider the DCI two-mass-easy box $I_4(2,3,5,6)$ depending on $\{x_2,x_3,x_5,x_6\}$. The Schubert problem has two solutions (Fig.~\ref{dci2me}) related by parity conjugation. As shown in the figure, while there are only two intersection points on $(AB)_1=(25)$, there are four on $(AB)_2=(\bar2\cap\bar5)$.

A line with four marked points gives cross-ratios forming an $A_1$ cluster algebra. Choose any two reference points, say $(12)\cap\bar5$ and $(23)\cap\bar5$. The four intersections on $(AB)_2$ can be written as linear combinations of the two:
\begin{equation*}
    \left(\begin{matrix}
        (12)\cap\bar5 & (23)\cap\bar5 & (45)\cap\bar2 & (56)\cap\bar2
    \end{matrix}\right)=\left(\begin{matrix}
        (12)\cap\bar5 & (23)\cap\bar5
    \end{matrix}\right)\times \left(\begin{matrix}
        1&0&-\langle2345\rangle&-\langle2356\rangle\\
        0&1&\langle1245\rangle&\langle1256\rangle
    \end{matrix}\right).
\end{equation*}
They form two cross-ratios which satisfy the $u$ equation of $A_1$:
\begin{equation}
    \mathcal U=\frac{(12)(34)}{(13)(24)}=\frac{\langle1235\rangle\langle2456\rangle}{\langle1245\rangle\langle2356},\quad\mathcal V=\frac{(14)(23)}{(13)(24)}=\frac{\langle1256\rangle\langle2345\rangle}{\langle1245\rangle\langle2356\rangle},\quad\mathcal U+\mathcal V=1.
\end{equation}

In general, a line with $m\geq4$ marked points is characterized by $m(m-3)/2$ cross-ratios, which form an $A_{m-3}$ cluster algebra. We now describe a canonical choice of these cross-ratios, the $u$-variables. Choose any two reference points $P,Q$ (which may or may not be marked) on the line, and write the $m$ marked points as their linear combinations:
\begin{equation*}
    \left(\begin{matrix}
        Z_1&\cdots&Z_m
    \end{matrix}\right)=\left(\begin{matrix}
        P&Q
    \end{matrix}\right)\times C_{2\times m}.
\end{equation*}
Due to the arbitrariness of $P,Q$ along the line and projectivity of twistors, the $C$ matrix actually lives in $G(2,m)/T$. The $m(m-3)/2$ cross-ratios can be chosen to be \[u_{ij}=\frac{(ij-1)(i-1j)}{(i-1j-1)(ij)},\quad\text{where }|i-j|\geq2.\]Here, $(ij)$ denotes the $(ij)$-minor of the $C$ matrix. These cross-ratios satisfy the $u$-equations \cite{Arkani-Hamed:2019plo}:
\begin{equation}
    u_{ij}+\prod_{(kl)\text{ crosses }(ij)}u_{kl}=1,
\end{equation}
making it clear that as $u_{ij}\to0$, all ``incompatible'' cross-ratios $u_{kl}$ with $i<k<j<l$ tend to 1.

Sometimes, we only need four out of the $m$ marked points. In this case, we simply ignore the other points and focus on the $A_1$ configuration. If the four points correspond to the $\{a,b,c,d\}$-th column of the $C$ matrix, the independent $A_1$ cross-ratios are simply $\mathcal U=\frac{(ab)(cd)}{(ac)(bd)}$ and $\mathcal V=\frac{(ad)(bd)}{(ac)(bd)}$. The extended $u$-equations \cite{Arkani-Hamed:2020tuz} express them as products of the $A_{m-3}$ $u$-variables:
\begin{equation}
    \mathcal U=\prod_{\substack{b\leq i<c\\d\leq j<a}}u_{ij},\quad\mathcal V=\prod_{\substack{a\leq i<b\\c\leq j<d}}u_{ij},\quad\mathcal U+\mathcal V=1.
\end{equation}
We remark that the \emph{logarithm} of such cross-ratios, namely the actual symbol letters, has a most natural integral representation associated with the geometry. Consider two intervals from any four points, {\it e.g.} $[a,d]$ and $[b,c]$, we integrate the unique $d\log$ form for one of them along the contour given by the other, which gives
\begin{equation}
    L([a,d],[b,c]):=\int_a^d{\rm d}\log \frac{(x c)}{(x b)}=\log\mathcal U=L([b,c],[a,d]),
\end{equation}
and similarly $L([a,b],[c,d])=L([c,d],[a,b])=\log\mathcal V$.

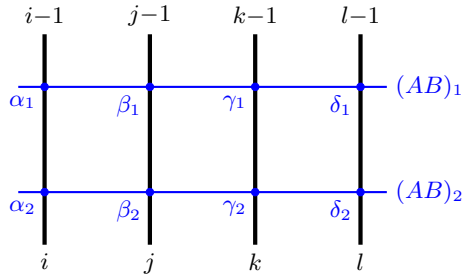
\begin{figure}[H]
    \centering
    \begin{tikzpicture}[scale=0.7]
        \draw[black,ultra thick](-4,2)--(-4,-2);
        \draw[black,ultra thick](-2,2)--(-2,-2);
        \draw[black,ultra thick](0,2)--(0,-2);
        \draw[black,ultra thick](2,2)--(2,-2);
        \draw[blue,thick](-4.5,1)--(2.5,1);
        \draw[blue,thick](-4.5,-1)--(2.5,-1);
        \filldraw[blue] (2.5,1) node[anchor=west] {{$(AB)_1$}};
        \filldraw[blue] (2.5,-1) node[anchor=west] {{$(AB)_2$}};
        \filldraw[black] (-4,2) node[anchor=south] {{$i{-}1$}};
        \filldraw[black] (-4,-2) node[anchor=north] {{$i$}};
        \filldraw[black] (-2,2) node[anchor=south] {{$j{-}1$}};
        \filldraw[black] (-2,-2) node[anchor=north] {{$j$}};
        \filldraw[black] (0,2) node[anchor=south] {{$k{-}1$}};
        \filldraw[black] (0,-2) node[anchor=north] {{$k$}};
        \filldraw[black] (2,2) node[anchor=south] {{$l{-}1$}};
        \filldraw[black] (2,-2) node[anchor=north] {{$l$}};
        \filldraw[blue]  (-4,1) circle [radius=2pt];
        \filldraw[blue]  (-4,-1) circle [radius=2pt];
        \filldraw[blue]  (-2,1) circle [radius=2pt];
        \filldraw[blue]  (-2,-1) circle [radius=2pt];
        \filldraw[blue]  (0,1) circle [radius=2pt];
        \filldraw[blue]  (0,-1) circle [radius=2pt];
        \filldraw[blue]  (2,1) circle [radius=2pt];
        \filldraw[blue]  (2,-1) circle [radius=2pt];
        \filldraw[blue] (-4,1) node[anchor=north east] {{$\alpha_1$}};
        \filldraw[blue] (-4,-1) node[anchor=north east] {{$\alpha_2$}};
        \filldraw[blue] (-2,1) node[anchor=north east] {{$\beta_1$}};
        \filldraw[blue] (-2,-1) node[anchor=north east] {{$\beta_2$}};
        \filldraw[blue] (0,1) node[anchor=north east] {{$\gamma_1$}};
        \filldraw[blue] (0,-1) node[anchor=north east] {{$\gamma_2$}};
        \filldraw[blue] (2,1) node[anchor=north east] {{$\delta_1$}};
        \filldraw[blue] (2,-1) node[anchor=north east] {{$\delta_2$}};
    \end{tikzpicture}
    \caption{Quadruple cut of the DCI four-mass box $I_4(i,j,k,l)$.}
    \label{dci4m}
\end{figure}
Let us spell out the details of the finite DCI four-mass box $I_4(i,j,k,l)$ (Fig.~\ref{dci4m}). Here, both solutions of the Schubert problem has four points, and remarkably the two pairs of cross-ratios~\cite{Yang:2022gko} read $\{z, 1-z\}$ and $\{\bar{z}, 1-\bar{z}\}$, satisfying $z \bar{z}={\bf u}=\frac{\langle i{-}1ij{-}1j\rangle \langle k{-}1kl{-}1l\rangle}{\langle i{-}1i k{-}1k\rangle\langle  j{-}1j l{-}1l\rangle}$, $(1-z)(1-\bar{z})={\bf v}=\frac{\langle i{-}1il{-}1l\rangle \langle j{-}1jk{-}1k\rangle}{\langle i{-}1i k{-}1k\rangle\langle  j{-}1j l{-}1l\rangle}$. They involve the square root $\Delta(x_i,x_j,x_k,x_l)=\sqrt{(1-\bf{u}-\bf{v})^2-4 {\bf u}{\bf v}}$, and famously these are the symbol letters of a large class of DCI integrals including the all-loop ladders~\cite{Drummond:2010cz}.

If we send $x_l$ to $I_\infty$, the DCI four-mass box becomes the non-DCI three-mass triangle $I_3(i,j,k)$. Two kinematics variables $\mathbf{u}$ and $\mathbf{v}$ become
\begin{equation}
  \mathbf{u}\to \frac{\langle i{-}1ij{-}1j\rangle \langle k{-}1kI_{\infty}\rangle}{\langle i{-}1i k{-}1k\rangle\langle  j{-}1j I_{\infty}\rangle}=\frac{m_1^2}{m_3^2},\ \mathbf{v}\to \frac{\langle j{-}1jk{-}1k\rangle \langle i{-}1iI_{\infty}\rangle}{\langle i{-}1i k{-}1k\rangle\langle  j{-}1j I_{\infty}\rangle}=\frac{m_2^2}{m_3^2}.
\end{equation}
Moreover, $\Delta(x_i,x_j,x_k,x_l)\to\frac{\Delta_{i,j,k}}{m_1^2m_2^2}$ with $\Delta_{i,j,k}{=}\sqrt{(m_3^2{-}m_1^2{-}m_2^2)^2{-}4 m_1^2m_2^2}$ reduces to the square root of the three-mass triangle, and $\{z,\bar z,1-z,1-\bar z\}$ becomes the three-mass triangle alphabet. As revealed in \cite{Chicherin:2020umh,He:2021eec,He:2022ctv}, it is a general fact that such non-DCI limits establish connections between DCI and non-DCI kinematics, integrals, and alphabets. On the other hand, since $I_\infty$ breaks dual conformal symmetry, the non-DCI kinematics has a higher dimension than the DCI kinematics. For instance, the four-mass box depends on two variables $\{\mathbf u,\mathbf v\}$ in the DCI case, while it depends on 5 dimensionless ratios of $\{m_1^2,m_2^2,m_3^2,m_4^2,s,t\}$ in the non-DCI case.

For future references, we explicitly present the two intersection points on $I_\infty$ of the three-mass triangle $I_3(i,j,k)$. Choose any two reference points $I_\infty=(PQ)$, the two intersection points read: (to save space, we have defined $i'\equiv i{-}1$)
\begin{equation}
    \alpha_{\pm}(ijk)=Z_P+Z_Q\frac12\frac{\langle Pkk^\prime(j^\prime j)\cap(ii^\prime Q)\rangle+\langle Qkk^\prime(j^\prime j)\cap(ii^\prime P)\rangle\pm\langle i^\prime ik^\prime k\rangle\langle j^\prime jPQ\rangle\Delta(x_i,x_j,x_k,I_\infty)}{\langle k^\prime k(j^\prime j)\cap(i^\prime iQ)Q\rangle}.
\end{equation}

\subsection{Boxes with up to 3 massive corners}

In this subsection and the next, we will look into some box and pentagon kinematics and obtain their one-loop alphabets by solving Schubert problems. In each case we will mainly consider intersections on $I_\infty$, together with possible $A_1$ configurations on the internal lines. We will see that cross-ratios from these $A_k\cup(A_1)^r$ configurations successfully reproduce their one-loop alphabets.

\paragraph{Zero-mass Box}To warm up, consider the zero-mass box with four one-mass triangle sub-topologies, each producing a pair of intersections on $I_\infty$, namely, $I_\infty\cap\overline{i-1}:=[i{-}2i{-}1i]$ and $I_\infty\cap\overline{i}:=[i{-}1ii{+}1]$. Hence, we obtain four distinct points on the line $I_\infty=(PQ)$, {\it i.e.}, an $A_1$ configuration. 
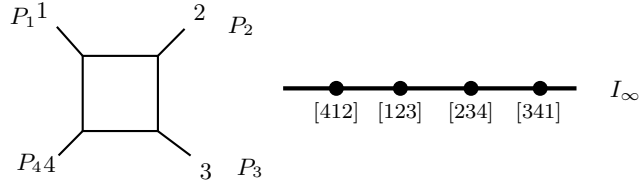
\begin{figure}[H]
    \centering
   \begin{tikzpicture}[baseline={([yshift=-4ex]current bounding box.center)},scale=0.2]
                \draw[black,thick] (0,5)--(-5,5)--(-5,0)--(0,0)--cycle;
                       \draw[black,thick] (0,5)--(1.7731,6.794);
                \draw[black,thick] (2.1728,-1.6351)--(-0.0479,0.0287);
                \draw[black,thick] (-6.7169,6.9345)--(-5,5);
                \draw[black,thick] (-5,0)--(-6.6028,-1.6224);
                \filldraw[black] (1.7871,6.8408) node[anchor=south west] {{$2$}};
                \filldraw[black] (2.2008,-1.6039) node[anchor=north west] {{$3$}};
                \filldraw[black] (-6.7112,6.9753) node[anchor=south east] {{$1$}};
                \filldraw[black] (-6.2212,-1.1446) node[anchor=north east] {{$4$}};
                   \filldraw[black] (-7.061,-0.9101) node[anchor=north east] {{$P_4$}};
                     \filldraw[black] (-7.4996,8.7485) node[anchor=north east] {{$P_1$}};
                       \filldraw[black] (6.9963,8.202) node[anchor=north east] {{$P_2$}};
                         \filldraw[black] (7.5868,-1.2632) node[anchor=north east] {{$P_3$}};
            \end{tikzpicture}
                        \begin{tikzpicture}[scale=1.3]
\draw [black, ultra thick](3,-5.2) -- (6.0,-5.2);
\node [fill=black,circle,inner sep=2pt] at (5.6286,-5.2) {};
\node [fill=black,circle,inner sep=2pt] at (4.9213,-5.2) {};
\node [fill=black,circle,inner sep=2pt] at (4.1989,-5.2) {};
\node [fill=black,circle,inner sep=2pt] at (3.5438,-5.2) {};
\node at (6.5,-5.22) {$I_\infty$};
\node [black] at (5.6138,-5.45) {\footnotesize $[341]$};
\node [black] at  (4.9189,-5.45) {{\footnotesize $[234]$}};
\node [black] at (4.1979,-5.45) {\footnotesize $[123]$};
\node [black] at (3.5446,-5.45) {\footnotesize $[412]$};
\end{tikzpicture}
    \caption{$A_1$ configuration for zero-mass box kinematics}
    \label{fig:0mbox}
\end{figure}
To compute the cross-ratios, we can parametrize the four points as follows:
\begin{equation*}
    \bordermatrix{ & [412] & [123] & [234] & [341] \cr
        Z_P & \langle Q412\rangle & \langle Q123\rangle & \langle Q234\rangle & \langle Q341\rangle \cr
        Z_Q & \langle412P\rangle & \langle123P\rangle & \langle234P\rangle & \langle341P\rangle
    }.
\end{equation*}
We obtain $u_{13}=\frac{\langle I_\infty12\rangle\langle I_\infty34\rangle}{\langle I_\infty13\rangle\langle I_\infty24\rangle}=\frac s{s+t}$ and $u_{24}=\frac{\langle I_\infty14\rangle\langle I_\infty23\rangle}{\langle I_\infty13\rangle\langle I_\infty24\rangle}=\frac t{s+t}$, which are precisely the multiplicatively-independent dimensionless ratios of symbol letters $\{s,t,s+t\}$. Notice that in this case, none of the Schubert solutions have four points on $(AB)$, so they do not produce non-trivial cross-ratios.

\paragraph{One-mass Box}For the one-mass box, the four triangle sub-topologies produce five distinct points on $I_\infty$, {\it  i.e.}, an $A_2$ configuration:
\begin{equation*}
    \bordermatrix{ & [123] & [124] & [234] & [452] & [345] \cr
        Z_P & \langle Q123\rangle & \langle Q124\rangle & \langle Q234\rangle & \langle Q452\rangle & \langle Q345\rangle \cr
        Z_Q & \langle123P\rangle & \langle124P\rangle & \langle234P\rangle & \langle452P\rangle & \langle345P\rangle
    }.
\end{equation*}
The five independent cross-ratios can be computed as
\begin{gather*}
    u_{13}=\frac{s}{m_1^2-t},\ u_{14}=\frac{t}{m_1^2-s},\ 
    u_{24}=\frac{m_1^2-s}{m_1^2},\ u_{35}=\frac{m_1^2-t}{m_1^2},\ 
    u_{25}=\frac{(m_1^2-s-t)m_1^2}{(m_1^2-s)(m_1^2-t)},
\end{gather*}
exactly matching the independent dimensionless ratios of the 6 Mandelstam letters:
\begin{equation*}
    \{m_1^2,\ s,\ t,\ m_1^2-s-t,\ m_1^2-s,\ m_1^2-t\}.
\end{equation*}
Note that although the one-mass box has two-mass triangle sub-topologies, all cross-ratios on their internal lines are contained in the above multiplicative space. For instance, for the Schubert solution $(AB)_1=((12)\cap\bar4,I_\infty\cap\bar4)$ of the two-mass triangle $I_3(2,4,5)$, the cross-ratios on $(AB)_1$ are $m_1^2/t$ and $(m_1^2-t)/t$.

\paragraph{Two-mass-easy Box}For the two-mass-easy box, there are six distinct points on $I_\infty$, {\it i.e.}, an $A_3$ configuration:
\begin{equation*}
    \bordermatrix{ & [123] & [125] & [235] & [452] & [562] & [456] \cr
        Z_P & \langle Q123\rangle & \langle Q125\rangle & \langle Q235\rangle & \langle Q452\rangle & \langle Q562\rangle & \langle Q456\rangle \cr
        Z_Q & \langle123P\rangle & \langle125P\rangle & \langle235P\rangle & \langle452P\rangle & \langle562P\rangle & \langle456P\rangle
    }.
\end{equation*}
The 9 independent cross-ratios exactly match the dimensionless ratios of 10 Mandelstam letters:
\[\{m_1^2,m_3^2,s,t,m_1^2-s,m_3^2-s,m_1^2-t,m_3^2-t,m_1^2+m_3^3-s-t,m_1^2m_3^2-s t\}.\]
Again, although some internal lines of triangle sub-topologies (and also the DCI Schubert solution of the two-mass-easy box itself) have four points, they do not produce any new letters.

\paragraph{Two-mass-hard Box} For the two-mass-hard box, a similar counting shows that there are six points on $I_\infty$, {\it i.e.}, an $A_3$ configuration with 9 independent cross-ratios.
\begin{figure}[H]
    \centering
    \begin{tikzpicture}[scale=0.25]
                \draw[black,thick] (0,5)--(-5,5)--(-5,0)--(0,0)--cycle;
                \draw[black,thick] (0,5)--(2.1651,5.6975);
                       \draw[black,thick] (0,5)--(0.6236,6.9805);
                \draw[black,thick] (1.9998,-1.6771)--(-0.0479,0.0287);
                \draw[black,thick] (-6.93,5.52)--(-5,5)--(-5.52,6.93);
                \draw[black,thick] (-5,0)--(-6.6028,-1.6224);
                \filldraw[black] (0.6376,7.0273) node[anchor=south west] {{$2$}};
                \filldraw[black] (2.1909,5.6727) node[anchor=west] {{$3$}};
                \filldraw[black] (2.0278,-1.6459) node[anchor=north] {{$4$}};
                \filldraw[black] (-6.93,6) node[anchor=east] {{$6$}};
                \filldraw[black] (-5.52,6.93) node[anchor=south] {{$1$}};
                \filldraw[black] (-6.6891,-1.6229) node[anchor=north east] {{$5$}};
                   \filldraw[black] (-7.3471,-0.8091) node[anchor=north east] {{$P_4$}};
                     \filldraw[black] (-6.0298,9.3106) node[anchor=north east] {{$P_1$}};
                       \filldraw[black] (5.4285,9.0369) node[anchor=north east] {{$P_2$}};
                         \filldraw[black] (5.6517,-0.7755) node[anchor=north east] {{$P_3$}};
            \end{tikzpicture}\quad\quad\begin{tikzpicture}[scale=1.5]
\draw [black, ultra thick](0.6504,-5.2) -- (6.1371,-5.2);
\draw [purple, ultra thick](0.392,-3.9631) -- (2.7432,-5.4902);
\node [fill=black,circle,inner sep=2pt] at (5.6286,-5.2) {};
\node [fill=black,circle,inner sep=2pt] at (4.8165,-5.1988) {};
\node [fill=black,circle,inner sep=2pt] at (4.1989,-5.2) {};
\node [fill=black,circle,inner sep=2pt] at (3.5438,-5.2) {};
\node [fill=black,circle,inner sep=2pt] at (2.9435,-5.2) {};
\node [fill=purple,circle,inner sep=2pt] at (2.3096,-5.2) {};
\node [fill=purple,circle,inner sep=2pt] at (1.0525,-4.4074) {};
\node [fill=purple,circle,inner sep=2pt] at (1.4396,-4.6346) {};
\node [fill=purple,circle,inner sep=2pt] at (1.8547,-4.9094) {};
\node at (6.4894,-5.2289) {$I_\infty$};
\node [black] at (5.6138,-5.4235) {\footnotesize $\alpha_+(246)$};
\node [black] at (4.1979,-5.4201) {{\footnotesize $[124]$}};
\node [black] at (2.9821,-5.4237) {\footnotesize$[125]$};
\node [purple] at (2.3016,-5.423) {\footnotesize$[345]$};
\node [purple] at (2.3072,-4.7922) {\footnotesize$(12)\cap\bar4$};
\node [purple] at (2.3689,-4.5342) {\footnotesize$(45)\cap(12(I_\infty\cap\bar4))$};
\node [purple] at (1.8683,-4.2632) {\footnotesize$(34)\cap(I_\infty((12)\cap\bar4))$};
\node [purple] at (0.3654,-3.8258) {\footnotesize$(I_\infty\cap\bar4,(12)\cap\bar4)$};
\node [black] at (3.5819,-5.431) {\footnotesize $[456]$};
\node [black] at (4.8141,-5.4174) {\footnotesize $\alpha_-(246)$};
\end{tikzpicture}
    \caption{$A_3\cup A_1$ for two-mass-hard box kinematics; $(I_\infty\cap\bar4,(12)\cap\bar4)$ is one of the Schubert solutions from $I_3(2,4,5)$}
    \label{fig:tmhbox}
\end{figure}
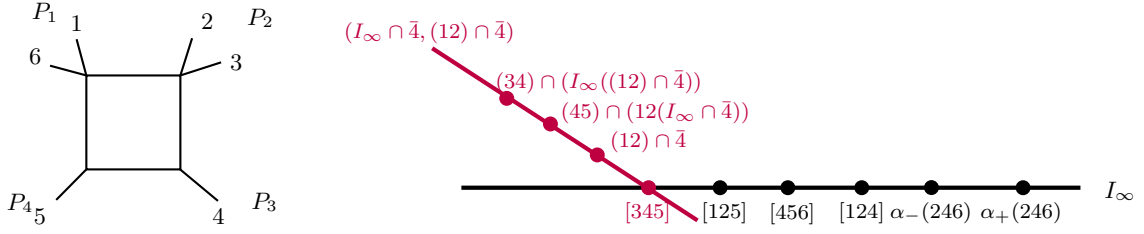
Note that due to the three-mass triangle sub-topology $I_3(2,4,6)$, we have $\alpha_\pm(246)$ on $I_\infty$, which contribute algebraic letters. However, the $A_3$ only spans a 9-dim subspace of the multiplicative space spanned by the 10 dimensionless ratios of the letters:
\[\{s,t,m_1^2,m_2^2,m_1^2-t,m_2^2-t,\Delta_{2,4,6},\frac{z}{\bar z},\frac{1-z}{1-\bar z},\frac{s z(1-\bar z)+t}{s\bar z(1-z)+t},m_1^2m_2^2-m_1^2t-m_2^2t+st+t^2\}.\]
For the first time, cross-ratios on the internal lines produce letters that cannot be obtained from those on $I_{\infty}$. In fact, any internal line with four points from $I_3(2,4,5)$, $I_3(2,5,6)$, or $I_3(2,4,6)$ suffices, and we have an $A_3\cup A_1$ configuration as above.

\paragraph{Three-mass Box}Finally, for the three-mass box, there are seven intersections on $I_\infty$, which constitutes an $A_4$ configuration with 14 independent cross-ratios.
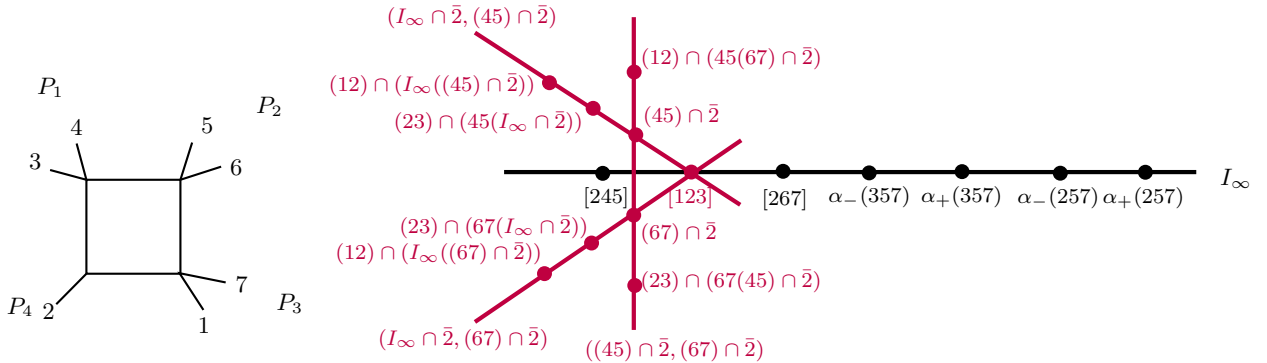
\begin{figure}[H]
    \centering
   \begin{tikzpicture}[baseline={([yshift=-16ex]current bounding box.center)},scale=0.25]
                \draw[black,thick] (0,5)--(-5,5)--(-5,0)--(0,0)--cycle;
                \draw[black,thick] (0,5)--(2.1651,5.6975);
                       \draw[black,thick] (0,5)--(0.6236,6.9805);
                \draw[black,thick] (1.2225,-1.9266)--(-0.0479,0.0287)--(2.4069,-0.4638);
                \draw[black,thick] (-6.93,5.52)--(-5,5)--(-5.52,6.93);
                \draw[black,thick] (-5,0)--(-6.6028,-1.6224);
                \filldraw[black] (0.6376,7.0273) node[anchor=south west] {{$5$}};
                \filldraw[black] (2.1909,5.6727) node[anchor=west] {{$6$}};
                \filldraw[black] (1.2505,-1.8954) node[anchor=north] {{$1$}};
                                \filldraw[black] (2.463,-0.5512) node[anchor=west] {{$7$}};
                \filldraw[black] (-6.93,6) node[anchor=east] {{$3$}};
                \filldraw[black] (-5.52,6.93) node[anchor=south] {{$4$}};
                \filldraw[black] (-6.2212,-1.1446) node[anchor=north east] {{$2$}};
                   \filldraw[black] (-7.3471,-0.8091) node[anchor=north east] {{$P_4$}};
                     \filldraw[black] (-5.7402,10.8914) node[anchor=north east] {{$P_1$}};
                       \filldraw[black] (5.9136,9.8645) node[anchor=north east] {{$P_2$}};
                         \filldraw[black] (6.9936,-0.6277) node[anchor=north east] {{$P_3$}};
            \end{tikzpicture}
             \begin{tikzpicture}[scale=1.5]
\draw [black, ultra thick](0.6504,-5.2) -- (6.785,-5.2);
\draw [purple, ultra thick](1.8,-3.8259) -- (1.8,-6.6);
\draw [purple, ultra thick](0.392,-3.9631) -- (2.7432,-5.4902);
\draw [purple, ultra thick](0.392,-6.5295) -- (2.7519,-4.918);
\node [fill=black,circle,inner sep=2pt] at (6.334,-5.2016) {};
\node [fill=black,circle,inner sep=2pt] at (5.58,-5.21) {};
\node [fill=black,circle,inner sep=2pt] at (4.7092,-5.196) {};
\node [fill=black,circle,inner sep=2pt] at (3.8862,-5.2072) {};
\node [fill=black,circle,inner sep=2pt] at (3.123,-5.1932) {};
\node [fill=black,circle,inner sep=2pt] at (1.5246,-5.2112) {};
\node [fill=purple,circle,inner sep=2pt] at (2.3096,-5.2) {};
\node [fill=purple,circle,inner sep=2pt] at (1.8087,-6.2058) {};
\node [fill=purple,circle,inner sep=2pt] at (1.8031,-4.3125) {};
\node [fill=purple,circle,inner sep=2pt] at (1.0525,-4.4074) {};
\node [fill=purple,circle,inner sep=2pt] at (1.4396,-4.6346) {};
\node [fill=purple,circle,inner sep=2pt] at (1.8175,-4.8696) {};
\node [fill=purple,circle,inner sep=2pt] at (1.0084,-6.1008) {};
\node [fill=purple,circle,inner sep=2pt] at (1.4256,-5.8294) {};
\node [fill=purple,circle,inner sep=2pt] at (1.7993,-5.582) {};
\node at (7.1373,-5.2569) {$I_\infty$};
\node [black] at (6.3192,-5.4251) {\footnotesize $\alpha_+(257)$};
\node [black] at (4.7082,-5.4161) {{\footnotesize $\alpha_+(357)$}};
\node [black] at (3.1384,-5.4419) {\footnotesize$[267]$};
\node [black] at (1.5561,-5.4295) {\footnotesize$[245]$};
\node [purple] at (2.3016,-5.423) {\footnotesize$[123]$};
\node [purple] at (2.2206,-4.7035) {\footnotesize$(45)\cap\bar2$};
\node [purple] at (0.5121,-4.7586) {\footnotesize$(23)\cap(45(I_\infty\cap\bar2))$};
\node [purple] at (0.0116,-4.4287) {\footnotesize$(12)\cap(I_\infty((45)\cap\bar2))$};
\node [purple] at (2.1979,-5.743) {\footnotesize$(67)\cap\bar2$};
\node [purple] at (0.5599,-5.6869) {\footnotesize$(23)\cap(67(I_\infty\cap\bar2))$};
\node [purple] at (0.0718,-5.9238) {\footnotesize$(12)\cap(I_\infty((67)\cap\bar2))$};
\node [purple] at (0.3654,-3.8258) {\footnotesize$(I_\infty\cap\bar2,(45)\cap\bar2)$};
\node [purple] at (0.2875,-6.6796) {\footnotesize$(I_\infty\cap\bar2,(67)\cap\bar2)$};
\node [black] at (3.8898,-5.4174) {\footnotesize $\alpha_-(357)$};
\node [black] at (5.5642,-5.4266) {\footnotesize $\alpha_-(257)$};
\node [purple] at (2.6636,-4.1933) {\footnotesize$(12)\cap(45(67)\cap\bar2)$};
\node [purple] at (2.6608,-6.1651) {\footnotesize$(23)\cap(67(45)\cap\bar2)$};
\node [purple] at (2.1612,-6.7777) {\footnotesize$((45)\cap\bar2,(67)\cap\bar2)$};
\end{tikzpicture}
    \caption{$A_4\cup(A_1)^3$ for three-mass box kinematics}
    \label{fig:3mbox}
\end{figure}
Moreover, we should go through possible internal lines from four triangle sub-topologies and the three-mass box itself. Together, they extend the dimension of the space spanned by cross-ratios by three. For example, in Fig.~\ref{fig:3mbox}, the three internal lines from $I_3(2,3,5)$, $I_3(2,3,7)$, and $I_4(2,3,5,7)$ shown in purple suffice.  In total, it forms an $A_4\cup(A_1)^3$ configuration and we obtain $17$ multiplicative independent cross-ratios from it, or $18$ letters in Mandelstam variables.  Only 17 of them actually appear in the one-loop CDE of the three-mass box. In the notation of \cite{Dlapa:2021qsl}, they are:
\begin{align*}
   \{W_1,\cdots, W_8, W_{11},W_{12},W_{15},\cdots W_{19}, W_{29},W_{30}\}.
\end{align*}The remaining one spanned by the cross-ratios can be identified as
\begin{equation}
    W_{23}=\frac{m_2^4-(m_1^2+m_3^2+s+t)m_2^2-(s-m_1^2)(t-m_3^2)+r_1r_3}{m_2^4-(m_1^2+m_3^2+s+t)m_2^2-(s-m_1^2)(t-m_3^2)-r_1r_3},
\end{equation}
where $r_1=\Delta_{2,5,7}$ and $r_3=\Delta_{3,5,7}$ are the square roots from three-mass triangle sub-topologies.
Although it does not appear as a one-loop letter, it is a letter of the \emph{two-loop} three-mass box.

\subsection{Pentagon kinematics and their one-loop alphabets}
\begin{figure}[H]
    \centering
  \begin{subfigure}{0.22\linewidth}
  \centering
    \begin{tikzpicture}[scale=0.2]
    \draw[black,thick](0,0)--(5,0)--(6.54,4.75)--(2.50,7.69)--(-1.54,4.75)--cycle;
    \draw[black,thick](-1.0717,-1.3182)--(0,0);
    \draw[black,thick](5.8442,-1.3436)--(5,0);
    \draw[black,thick](6.54,4.75)--(8.2859,5.4778);
    \draw[black,thick](-1.54,4.75)--(-3.2281,5.5542);
    \draw[black,thick](2.5,9.19)--(2.50,7.69);
    \filldraw[black] (-3.2281,5.5542) node[anchor=east] {{$5$}};
    \filldraw[black] (8.2859,5.4778) node[anchor=west] {{$2$}};
    \filldraw[black] (-1.0315,-1.3097) node[anchor=east] {{$4$}};
    \filldraw[black] (5.8028,-1.3988) node[anchor=west] {{$3$}};
    \filldraw[black] (1.7811,9.8619) node[anchor=west] {{$1$}};
    \filldraw[black] (1.0304,11.6582) node[anchor=west] {{$P_1$}};
    \filldraw[black] (7.1393,6.928) node[anchor=west] {{$P_2$}};
    \filldraw[black] (6.5961,-1.413) node[anchor=west] {{$P_3$}};
    \filldraw[black] (-5.8762,-1.1373) node[anchor=west] {{$P_4$}};
    \filldraw[black] (-6.1102,7.1036) node[anchor=west] {{$P_5$}};
    \end{tikzpicture}
    \caption{Zero-mass}
    \label{0mpenta}
    \end{subfigure}
    \begin{subfigure}{0.22\linewidth}
    \centering
    \begin{tikzpicture}[scale=0.2]
\draw[black,thick](0,0)--(5,0)--(6.54,4.75)--(2.50,7.69)--(-1.54,4.75)--cycle;
\draw[black,thick](-1.0717,-1.3182)--(0,0);
\draw[black,thick](5.8442,-1.3436)--(5,0);
\draw[black,thick](6.54,4.75)--(8.2859,5.4778);
\draw[black,thick](-1.54,4.75)--(-3.2281,5.5542);
\draw[black,thick](3.5,9)--(2.50,7.69)--(1.5,9);
\filldraw[black] (-3.2281,5.5542) node[anchor=east] {{$5$}};
\filldraw[black] (8.2859,5.4778) node[anchor=west] {{$2$}};
\filldraw[black] (-1.0315,-1.3097) node[anchor=east] {{$4$}};
\filldraw[black] (5.8028,-1.3988) node[anchor=west] {{$3$}};
\filldraw[black] (2.6084,9.671) node[anchor=west] {{$1$}};
\filldraw[black] (0.5,9.6726) node[anchor=west] {{$6$}};
\filldraw[black] (0.6144,11.3126) node[anchor=west] {{$P_1$}};
\filldraw[black] (7.1393,6.928) node[anchor=west] {{$P_2$}};
\filldraw[black] (6.5961,-1.413) node[anchor=west] {{$P_3$}};
\filldraw[black] (-5.8762,-1.1373) node[anchor=west] {{$P_4$}};
\filldraw[black] (-6.1102,7.1036) node[anchor=west] {{$P_5$}};
\end{tikzpicture}
\caption{One-mass}
\label{ompenta}
 \end{subfigure}
 \begin{subfigure}{0.22\linewidth}
 \centering
\begin{tikzpicture}[scale=0.2]
\draw[black,thick](0,0)--(5,0)--(6.54,4.75)--(2.50,7.69)--(-1.54,4.75)--cycle;
\draw[black,thick](-1.0717,-1.3182)--(0,0);
\draw[black,thick](6.0649,-0.9355)--(5,0);
\draw[black,thick](7.6952,6.3284)--(6.54,4.75)--(8.2521,4.6131);
\draw[black,thick](-1.54,4.75)--(-3.2281,5.5542);
\draw[black,thick](3.5,9)--(2.50,7.69)--(1.5,9);
\filldraw[black] (-3.2281,5.5542) node[anchor=east] {{$6$}};
\filldraw[black] (7.6707,6.2414) node[anchor=west] {{$2$}};
\filldraw[black] (-1.0315,-1.3097) node[anchor=east] {{$5$}};
\filldraw[black] (8.3568,4.6515) node[anchor=west] {{$3$}};
\filldraw[black] (6.1084,-1.0331) node[anchor=north] {{$4$}};
\filldraw[black] (2.6084,9.671) node[anchor=west] {{$1$}};
\filldraw[black] (0.5,9.6726) node[anchor=west] {{$7$}};
\filldraw[black] (0.6144,11.3126) node[anchor=west] {{$P_1$}};
\filldraw[black] (8.9445,6.0003) node[anchor=west] {{$P_2$}};
\filldraw[black] (6.9016,-1.4503) node[anchor=west] {{$P_3$}};
\filldraw[black] (-5.8762,-1.1373) node[anchor=west] {{$P_4$}};
\filldraw[black] (-6.1102,7.1036) node[anchor=west] {{$P_5$}};
\end{tikzpicture}
\caption{Two-mass-hard}
\label{tmhpenta}
 \end{subfigure}
 \begin{subfigure}{0.22\linewidth}
 \centering
\begin{tikzpicture}[scale=0.2]
\draw[black,thick](0,0)--(5,0)--(6.54,4.75)--(2.50,7.69)--(-1.54,4.75)--cycle;
\draw[black,thick](-1.0717,-1.3182)--(0,0);
\draw[black,thick](5.8442,-1.3436)--(5,0)--(6.9038,-0.3999);
\draw[black,thick](6.54,4.75)--(8.2859,5.4778);
\draw[black,thick](-1.54,4.75)--(-3.2281,5.5542);
\draw[black,thick](3.5,9)--(2.50,7.69)--(1.5,9);
\filldraw[black] (-3.2281,5.5542) node[anchor=east] {{$6$}};
\filldraw[black] (8.2859,5.4778) node[anchor=west] {{$2$}};
\filldraw[black] (-1.0315,-1.3097) node[anchor=east] {{$5$}};
\filldraw[black] (6.872,-0.5877) node[anchor=west] {{$3$}};
\filldraw[black] (5.8877,-1.4412) node[anchor=north] {{$4$}};
\filldraw[black] (2.6084,9.671) node[anchor=west] {{$1$}};
\filldraw[black] (0.5,9.6726) node[anchor=west] {{$7$}};
\filldraw[black] (0.6144,11.3126) node[anchor=west] {{$P_1$}};
\filldraw[black] (7.1393,6.928) node[anchor=west] {{$P_2$}};
\filldraw[black] (6.6809,-1.8584) node[anchor=west] {{$P_3$}};
\filldraw[black] (-5.8762,-1.1373) node[anchor=west] {{$P_4$}};
\filldraw[black] (-6.1102,7.1036) node[anchor=west] {{$P_5$}};
\end{tikzpicture}
\caption{Two-mass-easy}
\label{tmepenta}
 \end{subfigure}
\caption{Pentagon kinematics with no more than two massive corners}
\label{pentagons}
\end{figure}
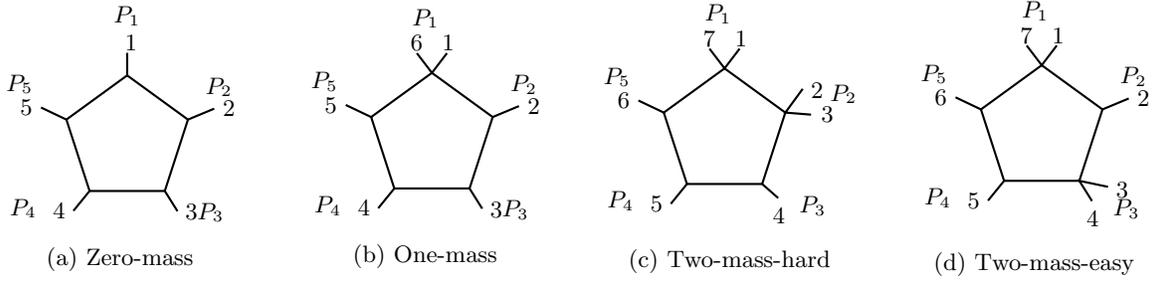

\paragraph{Zero-mass Pentagon}For the zero-mass pentagon (Fig.~\ref{0mpenta}), there are 10 intersection points on $I_\infty$ arising from its five one-mass box sub-topologies: $\{[512],\ [124],\ \text{+cyclic}\}$. Naively, one would expect an $A_7$ configuration with 35 independent cross-ratios. However, this cannot be true, since there are only four dimensionless kinematic variables (namely, ratios of $\{s_{12},s_{23},s_{34},s_{45},s_{51}\}$), and one needs at least seven to form an actual $A_7$. In fact, only 24 of the 35 cross-ratios are multiplicatively independent.

On the other hand, as emphasized in the main text, the 24-dim span of cross-ratios is not necessarily parity-invariant. Indeed, a multiplicative basis of the span can be chosen to be the dimensionless ratios of the following: (in the notation of \cite{Chicherin:2017dob})
\begin{equation}
    \begin{split}
        \Big\{W_1=s_{12},\ W_{11}=s_{12}-s_{45},\ W_{16}=s_{12}&+s_{23}-s_{45},\ W_{26}=\frac{s_{12}s_{23}-s_{23}s_{34}+s_{34}s_{45}-s_{45}s_{51}-s_{51}s_{12}+\mathop{\rm tr}_5}{s_{12}s_{23}-s_{23}s_{34}+s_{34}s_{45}-s_{45}s_{51}-s_{51}s_{12}-\mathop{\rm tr}_5},\\
        &-s_{12}s_{23}+s_{23}s_{34}+s_{34}s_{45}+s_{45}s_{51}+s_{51}s_{12}+{\mathop{\rm tr}}_5,\ \text{+cyclic}\Big\}.
    \end{split}
\end{equation}
We see that only the 19-dim span of dimensionless ratios of $\{W_1,W_{11},W_{16},W_{26},\text{+cyclic}\}$ is parity invariant. Together with $\mathop{\rm tr}_5$ itself, this alphabet precisely matches the one-loop symbol alphabet recorded in \cite{Chicherin:2017dob} .

All parity-even letters $\{W_1,W_{11},W_{16},\text{+cyclic}\}$ arise already from the union of five $A_2$ alphabets corresponding to the one-mass box sub-topologies. The parity-odd letters genuinely depend on five dual momenta, whose geometric origin is the mixing of triangle sub-topologies of different boxes. For example, the triangles $I_3(5,1,2)$ and $I_3(3,4,1)$ belong to different boxes, and their intersections on $I_\infty$
\[\{[456],[512],[234],[351]\}\]
produce $ \mathcal{U}^2=\frac{W_2W_{18}W_{30}}{W_3W_4W_{16}W_{27}},\ \mathcal{V}^2=\frac{W_5W_{11}^2W_{27}}{W_2W_{18}}$. It is interesting to observe that the product of numerator and denominator of $W_{26}$ is $4s_{45}s_{51}s_{12}(s_{51}-s_{23}-s_{34})=-4W_1W_4W_5W_{17}$, which belongs to parity-even sector of the alphabet.

\paragraph{One-mass Pentagon} For the one-mass pentagon (Fig.~\ref{ompenta}), there are 12 intersection points on $I_\infty$ arising from its five box sub-topologies. 40 of the 54 cross-ratios are multiplicatively independent. Moreover, since there are two-mass-hard box sub-topologies, it is crucial that we take into account the $A_1$ configurations on internal lines. All in all, we arrive at a 28 dimensional parity-invariant subspace of the span of $u$-variables. A basis can be chosen, in the notation of \cite{Abreu:2020jxa}, as the dimensionless ratios of
\begin{equation*}
    \{W_1,\cdots, W_9,\ W_{12},\cdots, W_{15},\ W_{18},W_{19},\ W_{22},\cdots,W_{24},\ W_{33},W_{34},\ W_{37},W_{38},\ W_{40},\ W_{43},\cdots, W_{48}\}.
\end{equation*}
Together with $W_{49}=\mathop{\rm tr}_5$ (Note that this ${\rm tr}_5$ is different from the one for zero-mass case.), we successfully reproduce the one-loop one-mass pentagon alphab

\paragraph{Two-mass-easy Pentagon}For the two-mass-easy pentagon (Fig.~\ref{tmepenta}), considering the 14 intersection points on $I_\infty$ and all possible internal lines, we get a 39 dimensional parity-invariant subspace. With $\mathop{\rm tr}_5$ itself, we obtain an alphabet of $41$ letters in Mandelstam variables. We find an almost perfect agreement with the CDE result (an alphabet of 40 letters) recorded in the ancillary file. The only discrepancy is that the geometric construction yields one more letter:
\begin{equation*}
    \frac{m_1^2m_3^2-m_1^2s_{12}-m_3^2s_{23}+s_{12}s_{23}+(m_1^2+m_3^2+s_{12}+s_{23})s_{45}-s_{45}^2+\Delta_1\Delta_2}{m_1^2m_3^2-m_1^2s_{12}-m_3^2s_{23}+s_{12}s_{23}+(m_1^2+m_3^2+s_{12}+s_{23})s_{45}-s_{45}^2-\Delta_1\Delta_2},
\end{equation*}
where $\Delta_1=\Delta_{2,5,7}=\sqrt{(m_1^2-s_{23}-s_{45})^2-4s_{23}s_{45}},\quad\Delta_2=\Delta_{3,5,7}=\sqrt{(m_3^2-s_{12}-s_{45})^2-4s_{12}s_{45}}$. This is nothing but the $W_{23}$ of the three-mass box sub-topology $I_4(2,3,5,7)$ at two loops.

\paragraph{Two-mass-hard Pentagon}For the two-mass-hard pentagon (Fig.~\ref{tmhpenta}), a new complication arises. As pointed out in \cite{He:2021non,He:2021esx}, the two-mass-hard pentagon kinematics has a non-trivial DCI sector, contributing a non-trivial DCI letter (in the notation of \cite{He:2021esx})
\begin{equation}
   \frac{\langle12\bar4\cap\bar6\rangle\langle1246\rangle}{\langle1245\rangle\langle1256\rangle\langle3467\rangle}=\frac{s_{51}s_{12}s_{23}-m_1^2s_{23}s_{34}-m_2^2s_{45}s_{51}}{s_{12}s_{23}s_{51}}
\end{equation} 
which by definition is not constructible on $I_\infty$. The solution is to construct cross-ratios on finite external lines, a lesson learned from \cite{Yang:2022gko}. For example, one way to obtain the factor $s_{12}s_{23}s_{51}-m_1^2s_{23}s_{34}-m_2^2s_{45}s_{51}$ is to mix the intersection points of $I_4(2,4,5,7)$ and $I_4(2,4,6,7)$ on the external line $(34)$.
\begin{equation}
    \bordermatrix{ & 3 & (67)\cap\bar4 & (34)\cap\bar1 & (34)\cap\bar6 \cr
        3 & 0 & \langle4(12)(35)(67)\rangle & -\langle1246\rangle & -\langle4567\rangle \cr
        4 & 1 & \langle3(12)(45)(67)\rangle & \langle1236\rangle & \langle3567\rangle
    },\quad\mathcal U=\frac{(m_1^2s_{34}-s_{51}s_{12})(m_2^2s_{45}-s_{12}s_{23})}{s_{12}(s_{51}s_{12}s_{23}-m_1^2s_{23}s_{34}-m_2^2s_{45}s_{51})}.
\end{equation}
In summary, we construct cross-ratios of three types: (a) intersections on $I_\infty$ from triangles; (b) intersections on internal lines; (c) intersections on finite external lines from boxes. The parity-invariant subspace of their span is 44 dimensional. Adding $\mathop{\rm tr}_5$, we obtain an alphabet of 46 letters in Mandelstam variables. The only discrepancy with the CDE result is the appearance of two additional two-loop letter: the $W_{23}$ for three-mass box sub-topologies $I_4(2,4,5,7)$ and $I_4(2,4,6,7)$.

\section{Details of two-loop Schubert problems and alphabets}
In this section, we present details about producing alphabets of two-loop integrals from Schubert problems. We will classify all two-loop symbol letters into two groups: letters from a single Schubert problem for a LS, and letters from combining different Schubert problems. The first group of letters are produced by taking (any cross-ratios containing) the minor of two intersections on $I_\infty$ from one Schubert problem, such as the two-loop rational letter $s{+}t$ for the one-mass slashed-box and the two-loop square root $\Delta_{nc}$ for the two-mass-easy slashed-box, which have been mentioned in the main text; they correspond to LS of individual integrals. Letters in the second group, on the other hand, only show up when we consider cross-ratios of four intersections on $I_\infty$ from different Schubert problems, which contains at least one minor involving two points from these two problems. The three algebraic letters \eqref{L3} of the two-mass-easy box kinematics belong to this group.

\subsection{One-mass pentagon kinematics and its two-loop alphabet}
Recall that for an individual two-loop diagram, by considering its LS, we may get new intersections on the infinity line $I_{\infty}:=(PQ)$. Such new points can form new cross-ratios (minors), which give us new letters. Let us first give some details about producing two-loop letters for the zero- and one-mass pentagon kinematics (Fig.~\ref{0mpenta},~\ref{ompenta}) from Schubert problems. The basic idea is the following. We first consider alphabets of their sub-topologies, and then those for genuinely 5-point two-loop diagrams, {\it i.e.}, integrals depending on all five dual momenta. For instance, the zero-mass pentagon has five one-mass box sub-topologies, each contributing one of the five two-loop letters $\{s_{12}{+}s_{23},\cdots,s_{15}+s_{12}\}$, while genuinely 5-point diagrams contribute no new letters. Therefore, we arrive at the two-loop zero-mass pentagon alphabet \cite{Chicherin:2017dob}. 

Now let us turn to the one-mass case. According to \cite{Abreu:2020jxa}, there are $49$ relevant and $9$ irrelevant letters in this case. Among them, $30$ letters are already produced from one-loop configurations, so we only need to generate the rest.

\paragraph{Two-loop letters from box sub-topologies}
Let us first look into its five box sub-topologies: two one-mass boxes $I_4(2,3,4,5)$ and $I_4(3,4,5,6)$, two two-mass-hard boxes $I_4(2,4,5,6)$ and $I_4(2,3,4,6)$, and one two-mass-easy box $I_4(2,3,5,6)$. Following the discussion of one-mass boxes, the two-loop letters $\{W_{20}=s_{23}{+}s_{34},W_{21}=s_{34}{+}s_{45}\}$ 
from $I_4(2,3,4,5)$ and $I_4(3,4,5,6)$ should show up in the final alphabet.

For the two-mass-easy box $I_4(2,3,5,6)$, as we mentioned in the main text, only the two-mass-easy slashed-box (Fig.~\ref{slashboxesb}) contains new letters, whose integrand reads
\begin{equation}
    \frac1{\langle AB12\rangle\langle AB56\rangle\langle ABCD\rangle\langle CD23\rangle\langle CD45\rangle\langle ABPQ\rangle\langle CDPQ\rangle}
\end{equation}
We parametrize $Z_A=\alpha_1 Z_1+\beta_1 Z_P+Z_Q$, $Z_B=\gamma_1 Z_1+\delta_1 Z_P+Z_2$ and $Z_C=\alpha_2 Z_2+\beta_2 Z_P+Z_Q$, $Z_D=\gamma_2 Z_2+\delta_2 Z_P+Z_3$. To compute its LS, we set all $7$ factors to $0$ and arrive at $\alpha_1=\alpha_2=0$, $\beta_1=\beta_2$, as well as a quadratic equation for $\beta_{1,2}$ from the vanishing of the composite LS Jacobian:
\begin{align}\label{tmee}
\langle Q,(23)\cap(45P),(56P)\cap(12P)\rangle\beta^2+(\langle P,(23)\cap(45P),(56Q)\cap(12Q)\rangle&-\langle Q,(23)\cap(45Q),(56P)\cap(12P)\rangle)\beta\nonumber\\
&+\langle P,(23)\cap(45Q),(56Q)\cap(12Q)\rangle=0
\end{align}
Its discriminant reads $\Delta_{nc}=\sqrt{(s+t)^2-4 m_1^2 m_3^2}$, which is a new square root (different from the one-loop triangle/box square root). We find two new intersections from this two-loop LS, which we denote as $\epsilon_\pm=\beta_\pm Z_P+Z_Q$ with $\beta_\pm$ being the two solutions of \eqref{tmee}), on $I_\infty$ (Fig.~\ref{box2mepq}). Thus, on top of the original (one-loop) $A_3$, we have an $A_5^\prime$ configuration (here the 8 points are not completely independent, {\it cf}. appendix VII.B). Computing all possible cross-ratios of $\epsilon_\pm$ with two arbitrary one-loop intersections, we reproduce a $3$-dimensional space spanned by algebraic letters containing this new square root, whose basis can be chosen to be \eqref{L3}. For instance, the $A_1$ configuration $\{[123],[245],\epsilon_+,\epsilon_-\}$ produces the cross-ratio $\mathcal V^{-2}=L_1 L_2$. In the notation of \cite{Abreu:2020jxa}, these $3$ algebraic letters correspond to $\{W_{35},W_{36},W_{39}\}$. Moreover, the LS itself is the irrelevant letter $W_{50}=\Delta_{nc}$.


Similarly, for the two-mass-hard box sub-topology, the LS of the following four two-loop integrals already give two-loop letters $\{m_1^2-s-t,m_2^2-s-t,m_1^2m_2^2-m_1^2t+s t,m_1^2 m_2^2-m_2^2t+s t\}$. Altogether such two two-mass-hard box sub-topologies account for $\{W_{11},W_{12},W_{16},W_{17},W_{27},W_{28},W_{29},W_{30}\}$ in the final alphabet.
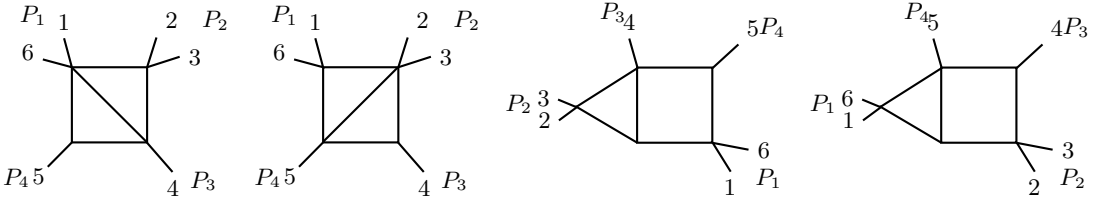
\begin{figure}[H]
    \centering
  \begin{tikzpicture}[scale=0.2]
                \draw[black,thick] (0,5)--(-5,5)--(-5,0)--(0,0)--cycle;
                \draw[black,thick] (0,5)--(2.1651,5.6975);
                       \draw[black,thick] (0,5)--(0.6236,6.9805);
                  \draw[black,thick] (-4.9981,4.9809)--(-0.0439,0.042);
                \draw[black,thick] (1.6908,-1.9674)--(-0.0479,0.0287);
                \draw[black,thick] (-6.93,5.52)--(-5,5)--(-5.52,6.93);
                \draw[black,thick] (-5,0)--(-6.6028,-1.6224);
                \filldraw[black] (0.6376,7.0273) node[anchor=south west] {{$2$}};
                \filldraw[black] (2.1909,5.6727) node[anchor=west] {{$3$}};
                \filldraw[black] (1.7188,-1.9362) node[anchor=north] {{$4$}};
                \filldraw[black] (-6.93,6) node[anchor=east] {{$6$}};
                \filldraw[black] (-5.52,6.93) node[anchor=south] {{$1$}};
                \filldraw[black] (-6.2212,-1.1446) node[anchor=north east] {{$5$}};
                   \filldraw[black] (-7.2126,-1.1439) node[anchor=north east] {{$P_4$}};
                  \filldraw[black] (-6.1069,9.6568) node[anchor=north east] {{$P_1$}};
                       \filldraw[black] (6.0154,9.3268) node[anchor=north east] {{$P_2$}};
                         \filldraw[black] (5.2037,-1.3457) node[anchor=north east] {{$P_3$}};
            \end{tikzpicture}
             \begin{tikzpicture}[scale=0.2]
                \draw[black,thick] (0,5)--(-5,5)--(-5,0)--(0,0)--cycle;
                \draw[black,thick] (0,5)--(2.1651,5.6975);
                       \draw[black,thick] (0,5)--(0.6236,6.9805);
                  \draw[black,thick] (-4.9981,0.042)--(-0.0439,4.9809);
                \draw[black,thick] (1.6908,-1.9674)--(-0.0479,0.0287);
                \draw[black,thick] (-6.93,5.52)--(-5,5)--(-5.52,6.93);
                \draw[black,thick] (-5,0)--(-6.6028,-1.6224);
                \filldraw[black] (0.6376,7.0273) node[anchor=south west] {{$2$}};
                \filldraw[black] (2.1909,5.6727) node[anchor=west] {{$3$}};
                \filldraw[black] (1.7188,-1.9362) node[anchor=north] {{$4$}};
                \filldraw[black] (-6.93,6) node[anchor=east] {{$6$}};
                \filldraw[black] (-5.52,6.93) node[anchor=south] {{$1$}};
                \filldraw[black] (-6.2212,-1.1446) node[anchor=north east] {{$5$}};
                   \filldraw[black] (-7.2126,-1.1439) node[anchor=north east] {{$P_4$}};
                     \filldraw[black] (-6.1069,9.6568) node[anchor=north east] {{$P_1$}};
                       \filldraw[black] (6.0154,9.3268) node[anchor=north east] {{$P_2$}};
                         \filldraw[black] (5.2037,-1.3457) node[anchor=north east] {{$P_3$}};
            \end{tikzpicture}
           \begin{tikzpicture}[scale=0.2]
                \draw[black,thick] (0,5)--(-5,5)--(-5,0)--(0,0)--cycle;
                \draw[black,thick] (-9.0156,2.3572)--(-10.1961,1.4845);
                       \draw[black,thick] (-9.0348,2.3668)--(-10.2534,2.8729);
                  \draw[black,thick] (-9.0673,2.3669)--(-4.9503,4.9789);
                  \draw[black,thick] (-9.1056,2.4339)--(-5.0077,0.0095);
                \draw[black,thick] (1.2225,-1.9266)--(0.0286,-0.0096)--(2.4069,-0.4638);
                \draw[black,thick] (-4.9713,4.9905)--(-5.52,6.93);
                \draw[black,thick] (-0.0005,4.9502)--(1.7346,6.538);
                \filldraw[black] (-10.2394,2.9197) node[anchor=east] {{$3$}};
                \filldraw[black] (-10.1703,1.4597) node[anchor=east] {{$2$}};
                \filldraw[black] (2.4582,-0.5223) node[anchor=west] {{$6$}};
                                \filldraw[black] (1.1763,-1.913) node[anchor=north] {{$1$}};
                \filldraw[black] (-5.52,6.93) node[anchor=south] {{$4$}};
                \filldraw[black] (1.7168,6.5389) node[anchor=south west] {{$5$}};
                   \filldraw[black] (2.5054,6.4661) node[anchor=south  west] {{$P_4$}};
                     \filldraw[black] (-5.1383,9.8604) node[anchor=north east] {{$P_3$}};
                       \filldraw[black] (-11.391,3.7927) node[anchor=north east] {{$P_2$}};
                         \filldraw[black] (5.2068,-1.1572) node[anchor=north east] {{$P_1$}};
            \end{tikzpicture}
            \begin{tikzpicture}[scale=0.2]
                \draw[black,thick] (0,5)--(-5,5)--(-5,0)--(0,0)--cycle;
                \draw[black,thick] (-9.0156,2.3572)--(-10.1961,1.4845);
                       \draw[black,thick] (-9.0348,2.3668)--(-10.2534,2.8729);
                  \draw[black,thick] (-9.0673,2.3669)--(-4.9503,4.9789);
                  \draw[black,thick] (-9.1056,2.4339)--(-5.0077,0.0095);
                \draw[black,thick] (1.2225,-1.9266)--(0.0286,-0.0096)--(2.4069,-0.4638);
                \draw[black,thick] (-4.9713,4.9905)--(-5.52,6.93);
                \draw[black,thick] (-0.0005,4.9502)--(1.7346,6.538);
                \filldraw[black] (-10.2394,2.9197) node[anchor=east] {{$6$}};
                \filldraw[black] (-10.1703,1.4597) node[anchor=east] {{$1$}};
                \filldraw[black] (2.4582,-0.5223) node[anchor=west] {{$3$}};
                                \filldraw[black] (1.1763,-1.913) node[anchor=north] {{$2$}};
                \filldraw[black] (-5.52,6.93) node[anchor=south] {{$5$}};
                \filldraw[black] (1.7168,6.5389) node[anchor=south west] {{$4$}};
                   \filldraw[black] (2.5054,6.4661) node[anchor=south  west] {{$P_3$}};
                     \filldraw[black] (-5.1383,9.8604) node[anchor=north east] {{$P_4$}};
                       \filldraw[black] (-11.391,3.7927) node[anchor=north east] {{$P_1$}};
                         \filldraw[black] (5.2068,-1.1572) node[anchor=north east] {{$P_2$}};
            \end{tikzpicture}
    \caption{Two loop integrals needed for two-mass-hard box kinematics}
\end{figure}

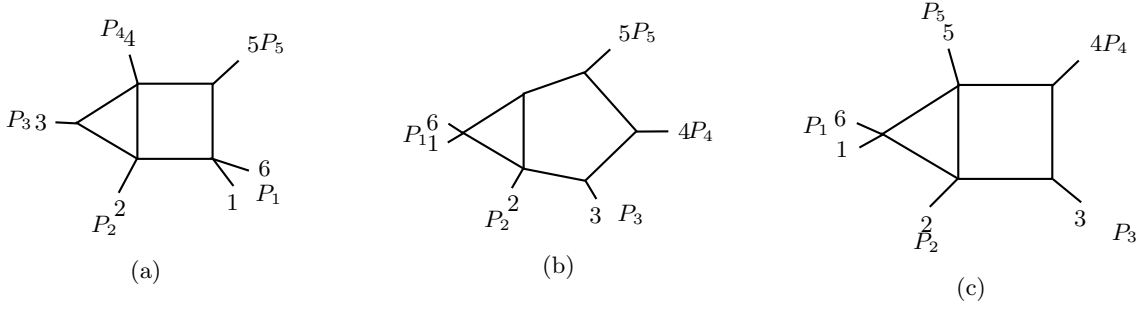
\begin{figure}[htbp]
    \centering
   \begin{subfigure}{0.30\linewidth}
   \begin{tikzpicture}[scale=0.2]
                \draw[black,thick] (0,5)--(-5,5)--(-5,0)--(0,0)--cycle;
                \draw[black,thick] (-9.0156,2.3572)--(-10.4471,2.4304);
                  \draw[black,thick] (-9.0673,2.3669)--(-4.9503,4.9789);
                  \draw[black,thick] (-9.1056,2.4339)--(-5.0077,0.0095);
                \draw[black,thick] (1.3854,-1.8045)--(0.0286,-0.0096)--(2.4157,-0.7853);
                \draw[black,thick] (-4.9713,4.9905)--(-5.52,6.93);
                \draw[black,thick] (-0.0005,4.9502)--(1.7346,6.538);
                                \draw[black,thick] (-5.0292,0.0096)--(-6.2431,-2.2594);
                \filldraw[black] (-10.4213,2.4056) node[anchor=east] {{$3$}};
                \filldraw[black] (-6.1485,-2.1897) node[anchor=north] {{$2$}};
                                \filldraw[black] (1.3392,-1.7909) node[anchor=north] {{$1$}};
                                                                \filldraw[black] (2.4845,-0.6564) node[anchor=west] {{$6$}};
                \filldraw[black] (-5.52,6.93) node[anchor=south] {{$4$}};
                \filldraw[black] (1.7168,6.5389) node[anchor=south west] {{$5$}};
                   \filldraw[black] (2.5054,6.4661) node[anchor=south  west] {{$P_5$}};
                     \filldraw[black] (-5.1383,9.8604) node[anchor=north east] {{$P_4$}};
                       \filldraw[black] (-11.391,3.7927) node[anchor=north east] {{$P_3$}};
                         \filldraw[black] (5.2068,-1.1572) node[anchor=north east] {{$P_1$}};
                                                  \filldraw[black] (-5.7282,-3.2139) node[anchor=north east] {{$P_2$}};
            \end{tikzpicture}
            \caption{}
            \label{boxtriangle1}
   \end{subfigure}
   \begin{subfigure}{0.30\linewidth}
    \begin{tikzpicture}[scale=0.2]
                \draw[black,thick] (-0.9774,6.4049)--(-5,5)--(-5,0)--(-0.9114,-0.8002)--(2.5,2.5)--cycle;
                \draw[black,thick] (-10,3)--(-9.0156,2.3572)--(-10.0425,1.7436);
                  \draw[black,thick] (-9.0673,2.3669)--(-4.9503,4.9789);
                  \draw[black,thick] (-9.1056,2.4339)--(-5.0077,0.0095);
                \draw[black,thick] (-0.1622,-2.0135)--(-0.8828,-0.8098);
                \draw[black,thick] (4.6196,2.4859)--(2.4827,2.4705);
                \draw[black,thick] (-0.9779,6.3551)--(0.7572,7.9429);
                                \draw[black,thick] (-5.0292,0.0096)--(-5.7816,-1.2898);
                \filldraw[black] (-10.0167,1.7188) node[anchor=east] {{$1$}};
                                \filldraw[black] (-10.0696,3.1092) node[anchor=east] {{$6$}};
                \filldraw[black] (-5.687,-1.2201) node[anchor=north] {{$2$}};
                                \filldraw[black] (-0.2084,-1.9999) node[anchor=north] {{$3$}};
                \filldraw[black] (4.7023,2.4951) node[anchor=west] {{$4$}};
                \filldraw[black] (0.7394,7.9438) node[anchor=south west] {{$5$}};
                   \filldraw[black] (1.528,7.871) node[anchor=south  west] {{$P_5$}};
                     \filldraw[black] (5.3284,2.4975) node[anchor=west] {{$P_4$}};
                       \filldraw[black] (-10.7131,3.4233) node[anchor=north east] {{$P_1$}};
                         \filldraw[black] (3.5805,-1.819) node[anchor=north east] {{$P_3$}};
                                                  \filldraw[black] (-5.2667,-2.2443) node[anchor=north east] {{$P_2$}};
            \end{tikzpicture}
            \caption{}
            \label{pentatri}
   \end{subfigure}
   \begin{subfigure}{0.30\linewidth}
    \begin{tikzpicture}[scale=0.25]
\draw[black,thick] (0,5)--(-5,5)--(-5,0)--(0,0)--cycle;
                \draw[black,thick] (-10.2572,1.6678)--(-9.0156,2.3572)--(-10.3955,2.9713);
                  \draw[black,thick] (-9.0673,2.3669)--(-4.9503,4.9789);
                  \draw[black,thick] (-9.1056,2.4339)--(-5.0077,0.0095);
                \draw[black,thick] (1.5379,-1.246)--(0.0286,-0.0096);
                \draw[black,thick] (-4.9713,4.9905)--(-5.52,6.93);
                \draw[black,thick] (1.4095,6.3017)--(-0.0005,4.9502);
                                \draw[black,thick] (-6.5152,-1.4864)--(-5.0292,0.0096);
                \filldraw[black] (-10.4824,3.2201) node[anchor=east] {{$6$}};
                                \filldraw[black] (-10.3943,1.5819) node[anchor=east] {{$1$}};
                                \filldraw[black] (-6.6912,-1.5561) node[anchor=north] {{$2$}};
                                \filldraw[black] (1.4917,-1.2324) node[anchor=north] {{$3$}};
                          \filldraw[black] (-5.52,6.93) node[anchor=south] {{$5$}};
                \filldraw[black] (1.5583,6.3362) node[anchor=south west] {{$4$}};
                   \filldraw[black] (2.1803,6.2298) node[anchor=south  west] {{$P_4$}};
                     \filldraw[black] (-5.1383,9.8604) node[anchor=north east] {{$P_5$}};
                       \filldraw[black] (-11.391,3.7927) node[anchor=north east] {{$P_1$}};
                         \filldraw[black] (5.0363,-1.9068) node[anchor=north east] {{$P_3$}};
                                                  \filldraw[black] (-5.5244,-2.4197) node[anchor=north east] {{$P_2$}};
            \end{tikzpicture}
            \caption{}
            \label{boxtriangle2}
   \end{subfigure}
    \caption{Genuine two-loop integrals needed for one-mass pentagon kinematics}
    \label{fig:my_label}
\end{figure}

\paragraph{Integrals with genuine one-mass pentagon kinematics}
Having considered the union of the alphabets from sub-topologies, only 6 relevant letters in \cite{Abreu:2020jxa} have not been accounted for: $\{W_{25},W_{26},W_{31},W_{32},W_{41},W_{42}\}$. Among them, $W_{32}$ is a letter corresponding to the LS of the integral in Fig.~\ref{boxtriangle1}. The intersections on $I_\infty$ are
\[\{X_1=-Z_P\frac{\langle5(12)(3Q)(46)\rangle}{\langle5(12)(3P)(46)\rangle}+Z_Q,X_2=[234]\}.\]
$W_{31}$, which is related to $W_{32}$ by reflection symmetry (relabeling the momentum twistors $\{1\leftrightarrow6,2\leftrightarrow5,3\leftrightarrow4\}$),  is  produced from its LS similarly.

Next we turn to the remaining four letters, which are letters from combining different Schubert problems and can be generated in many ways. For instance, in the case of Fig.~\ref{boxtriangle1}, we take all rational intersections together with $\{X_1\}$ on $I_\infty$ into account, and we get a configuration with $11$ points
\begin{align*}
    \{[123],[345],[456],[512],[523],[412],[356],[245],[256]\}\cup\{X_1,X_2=[234]\}
\end{align*}
Now we construct all independent cross-ratios with $\{X_1,X_2\}$. The conclusion is as follows. Up to multiplicative relations, there is only one independent parity-even factor:
\[\frac{(1,X_1)(4,X_2)}{(1,X_2)(4,X_1)}=\frac{W_3W_9}{W_{26}}\]
More explicitly, $W_{26}$ is produced by the minor $(X_1,[123])$. Moreover, although many cross-ratios involve ${\rm tr}_5$, there is only one combination that lives in the parity-invariant subspace, which produces the parity-odd letter $W_{42}$. The letters $W_{25}$ and $W_{41}$ are related by reflection, and can be produced similarly. We therefore obtain all relevant letters for one-mass pentagon kinematics up to two loops~\cite{Abreu:2020jxa}.

We can also produce irrelevant letters as follows. First of all, $W_{54}$ (an algebraic letter with square root $\Delta_{nc}$) can be generated by the $A_1$ configuration $\{[234],[345],\epsilon_+,\epsilon_-\}$ directly. Next, we  consider the penta-triangle integral in Fig.~\ref{pentatri}: by solving its LS, we get four pairs of solutions for $(AB), (CD)$, with {\it four} intersections on $I_\infty$:
\[\{X_1=-Z_P\frac{\langle2(1Q)(34)(56)\rangle}{\langle2(1P)(34)(56)\rangle}+Z_Q,\ X_2=[456],\ X_3=[356],\ X_4=[512]\}\]
Cross-ratios from these four points give $W_{53}$. Furthermore, by combining $X_1$ and one-loop intersections and doing a similar computation as $W_{42}$, we recover $W_{56}$. $W_{52}$ and $W_{55}$ can be generated by reflection similarly. Finally, for the remaining $W_{51}$, $W_{57}$ and $W_{58}$, we look into the box-triangle integral (Fig.~\ref{boxtriangle2}). We consider the Schubert problem where $(AB)$ intersects with $(23),(34),(45)$ and $I_\infty$, while $(CD)$ intersects with $(56),(12)$, $I_\infty$ and $(AB)$, and again we have four intersections on $I_{\infty}$:
\[\{X_1=[345],\ X_2=[234],\ X_3=Z_P+\frac{\langle P(12)(3Q)(56)\rangle}{\langle Q(12)(3P)(56)\rangle}Z_Q,\ X_4=Z_P+\frac{\langle P(12)(4Q)(56)\rangle}{\langle Q(12)(4P)(56)\rangle}Z_Q\}\,.\]
The cross-ratio $\frac{(X_1,X_2)(X_3,X_4)}{(X_1,X_3)(X_2,X_4)}=\frac{W_1W_2}{W_{51}}$ gives the even letter $W_{51}$, while cross-ratios of$\{X_3,X_4,\epsilon_+,\epsilon_-\}$ gives the odd letter $W_{58}$; similarly we recover the odd letter $W_{57}$ by combining $X_3,X_4$ with one-loop intersections.

\subsection{Three-mass box kinematics and its two-loop alphabet} The last case we present in this section is the two-loop alphabet for three-mass box kinematics. Following \cite{Dlapa:2021qsl}: there are $30$ letters in the alphabet for two-loop three-mass box-ladder, which after cyclic permutation yields all letters needed for three-mass box kinematics. Following the discussion in the last section, one-loop Schubert problems from three-mass box reproduce $18$ letters, and we need to produce the remaining $12$ letters (according to the notation in \cite{Dlapa:2021qsl}) 
\[\{W_9,W_{10},W_{13},W_{14},W_{20},W_{21},W_{22},W_{24},W_{25},W_{26},W_{27},W_{28}\}\,.\]
Let us look into two-loop diagrams with three-mass box kinematics. First of all, letters $\{W_9,W_{10},W_{13},W_{14}\}$
are from LS of the following four integrals respectively,
\begin{figure}[H]
    \centering
    \begin{tikzpicture}[scale=0.2]
                \draw[black,thick] (0,5)--(-5,5)--(-5,0)--(0,0)--cycle;
                \draw[black,thick] (0,5)--(2.1651,5.6975);
                       \draw[black,thick] (0,5)--(0.6236,6.9805);
                  \draw[black,thick] (-4.9981,4.9809)--(-0.0439,0.042);
                \draw[black,thick] (1.2225,-1.9266)--(-0.0479,0.0287)--(2.4069,-0.4638);
                \draw[black,thick] (-6.93,5.52)--(-5,5)--(-5.52,6.93);
                \draw[black,thick] (-5,0)--(-6.6028,-1.6224);
                \filldraw[black] (0.6376,7.0273) node[anchor=south west] {{$5$}};
                \filldraw[black] (2.1909,5.6727) node[anchor=west] {{$6$}};
                \filldraw[black] (1.2505,-1.8954) node[anchor=north] {{$1$}};
                                \filldraw[black] (2.463,-0.5512) node[anchor=west] {{$7$}};
                \filldraw[black] (-6.93,6) node[anchor=east] {{$3$}};
                \filldraw[black] (-5.52,6.93) node[anchor=south] {{$4$}};
                \filldraw[black] (-6.2212,-1.1446) node[anchor=north east] {{$2$}};
                   \filldraw[black] (-7.3471,-0.8091) node[anchor=north east] {{$P_4$}};
                     \filldraw[black] (-5.7402,10.8914) node[anchor=north east] {{$P_1$}};
                       \filldraw[black] (5.9136,9.8645) node[anchor=north east] {{$P_2$}};
                         \filldraw[black] (6.9936,-0.6277) node[anchor=north east] {{$P_3$}};
            \end{tikzpicture}
                \begin{tikzpicture}[scale=0.2]
                \draw[black,thick] (0,5)--(-5,5)--(-5,0)--(0,0)--cycle;
                \draw[black,thick] (-9.0156,2.3572)--(-10.1961,1.4845);
                       \draw[black,thick] (-9.0348,2.3668)--(-10.2534,2.8729);
                  \draw[black,thick] (-9.0673,2.3669)--(-4.9503,4.9789);
                  \draw[black,thick] (-9.1056,2.4339)--(-5.0077,0.0095);
                \draw[black,thick] (1.2225,-1.9266)--(0.0286,-0.0096)--(2.4069,-0.4638);
                \draw[black,thick] (-6.93,5.52)--(-4.9713,4.9905)--(-5.52,6.93);
                \draw[black,thick] (-0.0005,4.9502)--(1.7346,6.538);
                \filldraw[black] (-10.2394,2.9197) node[anchor=east] {{$5$}};
                \filldraw[black] (-10.1703,1.4597) node[anchor=east] {{$6$}};
                \filldraw[black] (2.4582,-0.5223) node[anchor=west] {{$1$}};
                                \filldraw[black] (1.1763,-1.913) node[anchor=north] {{$7$}};
                \filldraw[black] (-6.93,6) node[anchor=east] {{$4$}};
                \filldraw[black] (-5.52,6.93) node[anchor=south] {{$3$}};
                \filldraw[black] (1.7168,6.5389) node[anchor=south west] {{$2$}};
                   \filldraw[black] (2.5054,6.4661) node[anchor=south  west] {{$P_4$}};
                     \filldraw[black] (-5.7402,10.8914) node[anchor=north east] {{$P_1$}};
                       \filldraw[black] (-11.391,3.7927) node[anchor=north east] {{$P_2$}};
                         \filldraw[black] (5.2068,-1.1572) node[anchor=north east] {{$P_3$}};
            \end{tikzpicture}
                \begin{tikzpicture}[scale=0.2]
                \draw[black,thick] (0,5)--(-5,5)--(-5,0)--(0,0)--cycle;
                \draw[black,thick] (-9.0156,2.3572)--(-10.1961,1.4845);
                       \draw[black,thick] (-9.0348,2.3668)--(-10.2534,2.8729);
                  \draw[black,thick] (-9.0673,2.3669)--(-4.9503,4.9789);
                  \draw[black,thick] (-9.1056,2.4339)--(-5.0077,0.0095);
                \draw[black,thick] (1.2225,-1.9266)--(0.0286,-0.0096)--(2.4069,-0.4638);
                \draw[black,thick](-4.9713,4.9905)--(-6.6075,6.3064);
                \draw[black,thick] (-0.0005,4.9502)--(1.8716,5.8711);
                    \draw[black,thick] (-0.0005,4.9502)--(0.775,6.9517);
                \filldraw[black] (-10.2394,2.9197) node[anchor=east] {{$1$}};
                \filldraw[black] (-10.1703,1.4597) node[anchor=east] {{$7$}};
                \filldraw[black] (2.4582,-0.5223) node[anchor=west] {{$5$}};
                                \filldraw[black] (1.1763,-1.913) node[anchor=north] {{$6$}};
                \filldraw[black] (-6.6075,6.3064) node[anchor=south] {{$2$}};
                \filldraw[black] (0.8141,7.0779) node[anchor=south west] {{$3$}};
                            \filldraw[black] (1.8822,5.8607) node[anchor=west] {{$4$}};
                   \filldraw[black] (2.5054,6.4661) node[anchor=south  west] {{$P_1$}};
                     \filldraw[black] (-5.7402,10.8914) node[anchor=north east] {{$P_4$}};
                       \filldraw[black] (-11.391,3.7927) node[anchor=north east] {{$P_3$}};
                         \filldraw[black] (5.2068,-1.1572) node[anchor=north east] {{$P_2$}};
            \end{tikzpicture} 
            \begin{tikzpicture}[scale=0.2]
                \draw[black,thick] (0,5)--(-5,5)--(-5,0)--(0,0)--cycle;
                \draw[black,thick] (0,5)--(2.1651,5.6975);
                       \draw[black,thick] (0,5)--(0.6236,6.9805);
                  \draw[black,thick] (-4.9425,0.0619)--(-0.0181,5.0049);
                \draw[black,thick] (1.2225,-1.9266)--(-0.0246,-0.0062)--(2.4069,-0.4638);
                \draw[black,thick] (-6.93,5.52)--(-5,5)--(-5.52,6.93);
                \draw[black,thick] (-5,0)--(-6.6028,-1.6224);
                \filldraw[black] (0.6376,7.0273) node[anchor=south west] {{$5$}};
                \filldraw[black] (2.1909,5.6727) node[anchor=west] {{$6$}};
                \filldraw[black] (1.2505,-1.8954) node[anchor=north] {{$1$}};
                                \filldraw[black] (2.463,-0.5512) node[anchor=west] {{$7$}};
                \filldraw[black] (-6.93,6) node[anchor=east] {{$3$}};
                \filldraw[black] (-5.52,6.93) node[anchor=south] {{$4$}};
                \filldraw[black] (-6.2212,-1.1446) node[anchor=north east] {{$2$}};
                   \filldraw[black] (-7.3471,-0.8091) node[anchor=north east] {{$P_4$}};
                     \filldraw[black] (-5.7402,10.8914) node[anchor=north east] {{$P_1$}};
                       \filldraw[black] (5.9136,9.8645) node[anchor=north east] {{$P_2$}};
                         \filldraw[black] (6.9936,-0.6277) node[anchor=north east] {{$P_3$}};
            \end{tikzpicture}
    \caption{Two-loop integrals needed for three-mass box kinematics}
    \label{fig:my_label}
\end{figure}
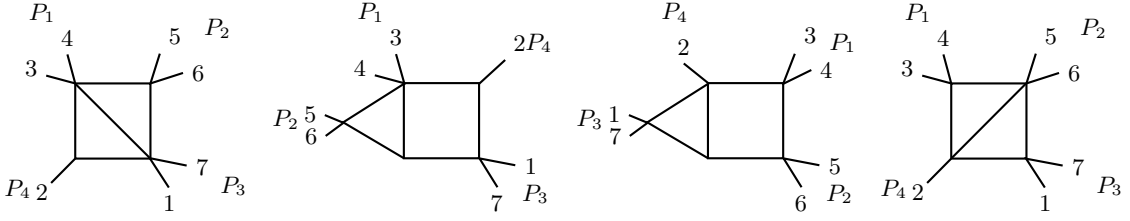
For the first two integrals, Schubert problem for each of them produces two intersections on $I_\infty$, one of which is new, and any cross-ratio containing the minor gives $W_9$ and $W_{10}$, respectively. For the last two cases LS of the integrals lead to square roots $\{\sqrt{W_{14}},\sqrt{W_{13}}\}$, and we have two new intersections on $I_{\infty}$ in each case (we denote the two points from the box-triangle as $\{X_1,X_2\}$, and $\{X_3,X_4\}$ for the slashed-box). 

The remaining $8$ letters $\{ W_{21},W_{26},W_{27},W_{28}; W_{20},W_{22},W_{24},W_{25}\}$ are from combining different Schubert problems. The first four letters involve $\sqrt{W_{14}}$, and the last four involve $\sqrt{W_{13}}$. For instance, to generate letters $\{W_{20},W_{22},W_{24},W_{25}\}$ involving $\sqrt{W_{13}}$, we take the following configuration with $9$ points into account,
\[\{[123],[245],[267],\alpha_+(2,5,7),\alpha_-(2,5,7),\alpha_+(3,5,7),\alpha_-(3,5,7),X_3,X_4\}\]
We combine $\{X_3,X_4\}$ (labelled as points $8,9$) with two one-loop intersections and compute cross-ratio correspondingly: \[\{\frac{(1,8)(2,9)}{(1,9)(2,8)}=W_{20}W_{22},\frac{(2,8)(3,9)}{(2,9)(3,8)}=W_{22},\frac{(6,8)(7,9)}{(6,9)(7,8)}=W_{24},\frac{(4,8)(5,9)}{(4,9)(5,8)}= W_{25}\}\]
thus recovering the four odd letters with $\sqrt{W_{13}}$.

Similarly, for those algebraic letters with $\sqrt{W_{14}}$, we consider the configuration with $9$ points
\[\{[123],[245],[267],\alpha_+(2,5,7),\alpha_-(2,5,7),\alpha_+(3,5,7),\alpha_-(3,5,7),X_1,X_2\}\]
whose cross-ratios involving $X_1, X_2$ give
\[\{\frac{(1,8)(3,9)}{(1,9)(3,8)}=\frac{W_{21}}{W_{28}},\frac{(4,8)(5,9)}{(4,9)(5,8)}=W_{26},\frac{(6,8)(7,9)}{(6,9)(7,8)}=W_{27},(\frac{(1,8)(5,9)}{(1,9)(5,8)})^2=\frac{W_{21}^2W_{26}}{W_{28}}\}\]
thus recovering all two-loop odd letters. Note that in both cases, we may consider other cross-ratios and obtain other odd letters with the two square roots. However, products of their numerators and denominators consist of factors that are not rational letters we have, which exclude them as valid symbol letters.



\section{Details of CDE for one-loop integrals}
Using integration-by-parts identities (IBPs)~\cite{CHETYRKIN1981159}, we can express any Feynman integral of a particular topology by a linear combination of the so-called master integrals (MIs) $ \vec{I} $, which form a basis for the concerned integral family. Derivatives of MIs can be reduced to this basis as well. Then differential equations (DE) for MIs can be obtained as: $ d \vec{I}( \vec{s},\epsilon )= d \mathbf{A}( \vec{s},\epsilon ) \vec{I}( \vec{s},\epsilon ) $, with $ d\mathbf{A} $ the derivative matrix, $ \vec{s} $ the kinematic variables. 
It is possible that, after a proper transformation of MIs, the DE can be brought into the so-called canonical differential equation (CDE) form~\cite{Henn:2013pwa,Henn:2014qga} 
\begin{equation}\label{key}
	d \vec{g}( \vec{s},\epsilon ) = \epsilon \, d \tilde{\mathbf{A}}( \vec{s} ) \vec{g}
	= \epsilon \, \sum_i \mathbf{C}_i d \log W_i \,  \vec{g}( \vec{s},\epsilon ) \, ,
\end{equation}
where the dependence on $ \epsilon $ factorizes and $  d \tilde{\mathbf{A}} $ takes the $ d\log $-form, with a summation over all letters $ W_i $ in the alphabet, and constant matrices $ \mathbf{C}_i $. The integrals in the properly-transformed basis $ \vec{g}( \vec{s},\epsilon ) $ are known to have uniform transcendentality~\cite{Henn:2013pwa}, hence we call $ \vec{g}( \vec{s},\epsilon ) $ the uniform transcendental (UT) basis. 
In the rest of this appendix, we review these bases for one-loop box integrals with one and two masses (note the subscripts of the integrals represent propagators with momenta $ \ell $, $(\ell-P_1)$, $(\ell-P_1-P_2)$, $ (\ell-P_1-P_2-P_3)$ respectively.), and for simplicity only the top sector of the system, {\it i.e.} the first line of the CDE system. We use these well-known examples to show that the alphabets from CDE indeed coincide with the space spanned by intersections of one-loop Schubert problems. We also attach an ancillary file which contains the CDE and alphabets for one-loop integrals with two-mass-esay and two-mass-hard pentagon kinematics. 
 
\paragraph{One-mass box}
Recall our kinematic variables $ s=(P_1+P_2)^2 $ and $ t=(P_2+P_3)^2 $, and $m_1^2=P_1^2$.  The UT basis reads:
\begin{equation}\label{key}
\begin{aligned}
&g_1 = \frac{1}{4} m_1^{2\epsilon} s t \epsilon ^2 \, \text{I}_{1,1,1,1}\, ,
&g_2 = m_1^{2\epsilon} (1-2 \epsilon ) \epsilon \, \text{I}_{1,1,0,0}\, , \\
&g_3 = m_1^{2\epsilon} (1-2 \epsilon ) \epsilon \, \text{I}_{1,0,1,0}\, , 
&g_4 = m_1^{2\epsilon} (1-2 \epsilon ) \epsilon \, \text{I}_{0,1,0,1}\, .
\end{aligned}	
\end{equation}
The top sector CDE reads:
\begin{equation}\label{key}
	\begin{aligned}
		d g_1 = 
		\epsilon \left[d \log\left( \frac{m_1^2-s-t}{s t} \right) g_1
		+ \frac{1}{2} d \log \left( \frac{m_1^2 (m_1^2-s-t)}{ (m_1^2-s)(m_1^2-t)} \right) g_2
		+ \frac{1}{2} d \log \left( \frac{m_1^2-s }{m_1^2-s-t} \right) g_3 \right. \\
		\left. + \frac{1}{2} d \log \left( \frac{m_1^2-t }{m_1^2-s-t} \right) g_4 \right]
	\end{aligned}
\end{equation}
which contains all the letters, agreeing with the $A_2$ alphabet. 

\paragraph{Two-mass-easy box} The UT basis for two-mass-easy box integral includes:
\begin{equation}\label{key}
	\begin{aligned}
		& g_1 = \frac{1}{4} m_1^{2\epsilon}  \left(s t-m_1^2 m_3^2\right) \epsilon ^2 \, \text{I}_{1,1,1,1} \, , 
		& g_2 = m_1^{2\epsilon} (1-2 \epsilon ) \epsilon \, \text{I}_{1,1,0,0}\, , 
		& \quad  g_3 = m_1^{2\epsilon} (1-2 \epsilon ) \epsilon \, \text{I}_{1,0,1,0}\, , \\
		& g_4 = m_1^{2\epsilon} (1-2 \epsilon ) \epsilon \, \text{I}_{0,1,0,1}\, , 
		& g_5 = m_1^{2\epsilon} (1-2 \epsilon ) \epsilon \, \text{I}_{0,0,1,1}\, .\\
	\end{aligned}
\end{equation}
where $ m_1^2=P_1^2, m_3^2=P_3^2 $. And the top sector CDE reads:
\begin{equation}\label{key}
	\begin{aligned}
		d g_1 = 
		\epsilon \left[	d \log\left( \frac{m_1^2+m_3^2-s-t}{m_1^2 m_3^2-st} \right) g_1 \right.
		+\frac{1}{2} d \log \left( \frac{m_1^2 (m_1^2+m_3^2-s-t)}{(m_1^2-s)(m_1^2-t)} \right) g_2
		+ \frac{1}{2} d \log \left( \frac{(m_1^2-s)(m_3^2-s)}{s(m_1^2+m_3^2-s-t)} \right) g_3\\
		+ \frac{1}{2} d \log \left( \frac{(m_1^2-t)(m_3^2-t)}{t(m_1^2+m_3^2-s-t)} \right) g_4
		\left. + \frac{1}{2} d \log \left( \frac{m_3^2 (m_1^2+m_3^2-s-t)}{(m_3^2-s)(m_3^2-t)}  \right) g_5 \right]
	\end{aligned}
\end{equation}
which nicely contain all letters of our $A_3$ alphabet. 
\paragraph{Two-mass-hard box} The UT basis for two-mass-hard box is
\begin{equation}\label{key}
	\begin{aligned}
		& g_1 = \frac{1}{4} m_1^{2\epsilon} s t \epsilon ^2 \, \text{I}_{1,1,1,1}\, ,
		&& g_2 = \frac{1}{2} m_1^{2\epsilon}  \epsilon ^2  \sqrt{(m_1^2+m_2^2-s)^2-4 m_1^2 m_2^2}\, \text{I}_{1,1,1,0} \, ,\\
		& g_3 =m_1^{2\epsilon} (1-2 \epsilon ) \epsilon \, \text{I}_{1,1,0,0}\, ,
		&& g_4 =m_1^{2\epsilon} (1-2 \epsilon ) \epsilon \, \text{I}_{1,0,1,0}\, ,\\
		& g_5 = m_1^{2\epsilon}(1-2 \epsilon ) \epsilon \, \text{I}_{0,1,1,0}\, , 
		&& g_6 = m_1^{2\epsilon}(1-2 \epsilon ) \epsilon \, \text{I}_{0,1,0,1}\, . \\
	\end{aligned}
\end{equation}
where $ m_1^2=P_1^2, m_2^2=P_2^2 $. The top sector CDE reads:
\begin{equation}\label{key}
	\begin{aligned}
		d g_1 =& 
		\epsilon \left[	d \log\left( \frac{W_7}{W_1 W_2^2} \right) g_1 \right.
		-\frac{1}{2} d \log \left( W_9 \right) g_2
		+ \frac{1}{4} d \log \left( \frac{W_4 W_7}{W_3^2 W_6} \right) g_3\\
		&+ \frac{1}{4} d \log \left( \frac{W_4 W_6}{W_7} \right) g_4
		\left. + \frac{1}{4} d \log \left( \frac{W_6 W_7}{W_4 W_5^2}  \right) g_5
		+ \frac{1}{2} d \log \left( \frac{W_3 W_5}{W_7}  \right) g_6 \right]
	\end{aligned}
\end{equation}
and we also need other equations (omitted here for simplicity) to see all the letters of the $A_3 \cup A_1$ alphabet, which are defined as
\begin{equation}\label{key}
	\begin{aligned}
	& \{W_1, W_2,\cdots,W_{11}\}=\\
	\{& m_1^2,s,t,t-m_1^2,m_2^2,t-m_2^2, t \left(-m_2^2+s+t\right)+m_1^2 \left(m_2^2-t\right),4 m_1^2 m_2^2-\left(m_1^2+m_2^2-s\right){}^2,\\
	& \frac{m_1^2-m_2^2+s-\sqrt{(m_1^2+m_2^2-s)^2-4 m_1^2 m_2^2}}{m_1^2-m_2^2+s+\sqrt{(m_1^2+m_2^2-s)^2-4 m_1^2 m_2^2}},\\
	& \frac{m_1^2+m_2^2-s-\sqrt{(m_1^2+m_2^2-s)^2-4 m_1^2 m_2^2}}{m_1^2+m_2^2-s+\sqrt{(m_1^2+m_2^2-s)^2-4 m_1^2 m_2^2}},\\
	&\frac{m_1^2 \left(t-2 m_2^2\right)+m_2^2 t-s t-t \sqrt{(m_1^2+m_2^2-s)^2-4 m_1^2 m_2^2}}{m_1^2 \left(t-2 m_2^2\right)+m_2^2 t-s t+t \sqrt{(m_1^2+m_2^2-s)^2-4 m_1^2 m_2^2}}\}
	\end{aligned}
\end{equation}

\paragraph{Two-mass pentagons} The results of two-mass pentagons are lengthy, so we put them in the ancillary file, where the UT bases $\vec{g}$ and the derivative matrices $ d \mathbf{\tilde{A}} $ in canonical form are given. 

\end{document}